\documentclass[12pt,letterpaper]{article} 
\usepackage[left = 2.5cm, right = 2.5cm, top = 2.5cm, bottom = 2.5cm]{geometry} 
\usepackage{setspace}
\usepackage{graphicx}
\usepackage{amsfonts} 
\usepackage{amsmath}
\usepackage{float}
\restylefloat{table}
\usepackage{natbib} 
\usepackage{placeins}
\usepackage{caption}
\usepackage{subcaption} 
\usepackage{amsthm}
\numberwithin{equation}{section}
\usepackage[ruled]{algorithm2e}
\usepackage{titlepic}
\DeclareMathAlphabet{\mathpzc}{OT1}{pzc}{m}{it}

\usepackage{color}
\usepackage[toc,page]{appendix}
\usepackage{fancyhdr} 
\usepackage{url}
\usepackage{xcolor}
\usepackage{amsmath}
\usepackage{authblk}

\newtheorem{theorem}{Theorem}[section]

\newtheorem{lemma}[theorem]{Lemma}

\usepackage{color}

\begin{document} 

\def\spacingset#1{\renewcommand{\baselinestretch}%
{#1}\small\normalsize} \spacingset{1}

\title{A computationally efficient nonparametric approach for changepoint detection}

\author[1$\dag$]{Kaylea Haynes}
\author[2]{Paul Fearnhead}
\author[3]{Idris A. Eckley}
\affil[1]{STOR-i Centre for Doctoral Training, Lancaster University}
\affil[2]{Department of Mathematics and Statistics, Lancaster University}
\affil[$\dag$]{Correspondence: k.haynes1@lancaster.ac.uk}
\date{}
\renewcommand\Authands{ and }
\spacingset{1.45}
\maketitle

\begin{abstract} 
In this paper we build on an approach proposed by \cite{Zou2014} for nonparametric changepoint detection. This approach defines the best segmentation for a data set as the one which
minimises a penalised cost function, with the cost function defined in term of minus a non-parametric log-likelihood for data within each segment. 
Minimising this cost function is possible using dynamic programming, but their algorithm had a computational cost that is cubic in the length of the data set.  
To speed up computation, \cite{Zou2014} resorted to a screening procedure which means that the estimated segmentation is no longer 
guaranteed to be the global minimum of the cost function. We show that the screening procedure adversely affects the accuracy of the changepoint
detection method, and show how a  faster dynamic programming algorithm, Pruned Exact Linear Time, PELT \citep{Killick2012}, can be used to
find the optimal segmentation with a computational cost that can be close to linear in the amount of data.  
PELT requires a penalty to avoid under/over-fitting the model which can have a detrimental effect on the quality of the detected changepoints. 
To overcome this issue we use a relatively new method, Changepoints Over a Range of PenaltieS (CROPS) \citep{Haynes14}, which finds all of the optimal segmentations 
for multiple penalty values over a continuous range. We apply our method to detect changes in heart rate during physical activity.    
\end{abstract}


\section{Introduction}
\label{Motivation}
Changepoint detection is an area of statistics broadly studied across many disciplines such as acoustics \citep{Guarnaccia2015,Lu2002}, genomics \citep{Olshen2004,Zhang2007b} and oceanography \citep{Nam2014}.  
Whilst the changepoint literature is vast, many existing methods are parametric. For example a common approach 
is to introduce a model for the data within a segment, use minus the maximum of the resulting log-likelihood to define a cost for a segment, and then define a cost of a segmentation as the sum of the costs for
each of its segments.  See for example \cite{Yao1988, Lavielle2005,Killick2012,Davis2006}. Finally, the segmentation of the data is obtained as the one that minimises a penalised version of this cost 
\cite[see also][for an extension of these approaches]{frick2014multiscale}.

A second class of methods are based on tests for a single changepoint, with the tests often defined based on the type of change that is expected (such as change in mean), and the distribution of the null-statistic
for each test depending on further modelling assumptions for the data \cite[see e.g.][]{Bai1998,DetteWied2015}.  Tests for detecting a single change can then be applied recursively to detect multiple changes, for example using binary segmentation 
\cite[]{Scott1974} or its variants \cite[e.g.][]{WildBS}. For a review of alternative approaches for change detection see \cite{Jandhyala2013} 
and \cite{Aue2013}.


Much of the existing literature on nonparametric methods look at single changepoint detection \citep{Page1954, Bhattacharyya1968, Carlstein1988, Dumbgen1991}.  
Several approaches are based on using rank statistics such as the Mann-Whitney test statistic \citep{Pettitt1979}.  \cite{Ross2012} introduce the idea of using the 
Kolmogorov-Smirnov and the Cramer-von Mises test statistics; both of which use the empirical distribution function.  Other methods include using kernel density estimations \citep{Baron2000}, 
however these can be computationally expensive to calculate.   

There is less literature on the nonparametric multiple changepoint setting. The single changepoint detection methods which have been developed using nonparametric methods do not extend easily to multiple changepoints. 
Within the sequential changepoint detection literature one can treat the problem as a single changepoint problem which resets every time a changepoint is detected \citep{Ross2012}.  \cite{Lee1996295} proposed a weighted empirical measure which is simple to use but has been shown to have unsatisfactory results.  Under the multivariate setting \cite{Matteson2013} proposed methods, E-divisive and e-cp3o, based on clustering and probabilistic pruning respectively. The E-divisive method uses an exact test statistic with an approximate search algorithm whereas the e-cp3o method uses an approximate test statistic with an exact search algorithm.  As a result e-cp3o is faster but lacks slightly in the quality for the changepoints detected.      

In this article we focus on univariate changepoint detection and we are interested in the work of \cite{Zou2014} who propose a nonparametric likelihood based on the empirical distribution.  
They then use a dynamic programming approach, Segment Neighbourhood Search \citep{IVANE.AUGERandCHARLESE.LAWRENCE1989}, which is an exact search procedure, to find multiple changepoints.  Whilst this method is shown to perform well, it has a computational 
cost of $\mathcal{O}(Mn^2+n^3)$ where 
$M$ is the number of changepoints and $n$ is the length of the data.  This makes this method infeasible when we have large data sets, particularly in situations where the number of changepoints increases with $n$.  To overcome
this, \cite{Zou2014} propose an additional screening step that prunes many possible changepoint locations. However, as we establish in this article,  this screening step can adversely affect the accuracy of the final inferred segmentation.

In this paper we seek to develop a computationally efficient approach to the multiple changepoint search problem in the nonparametric setting.   
Our approach is an extension to the method of \cite{Zou2014}. This firstly involves simplifying the definition of the segment cost, so that calculating the cost for a given segment involves
computation that is $O(\log n)$ rather than $O(n)$. Secondly we apply a different dynamic programming approach, Pruned Exact Linear Time (PELT) \citep{Killick2012}, that is substantially quicker than Segment Neighbourhood Search; for many
situations where the number of changepoints increases linearly with $n$, PELT has been proven to have a computational cost that is linear in $n$. 

We call the new algorithm nonparametric PELT (NP-PELT). A disadvantage
of NP-PELT is that it requires the user to pre-specify a value by which the addition of a changepoint is penalised. The quality of the final segmentation can be sensitive to this choice, and whilst there are default choices these do
not always work well. However we show that the Changepoints  for a Range of PenaltieS (CROPS) algorithm \citep{Haynes14} can be used with NP-PELT to explore optimal segmentations for a range of penalties.

The rest of this paper is organised as follows.  In Section \ref{background} we give details of the NMCD approach proposed by \cite{Zou2014}.  In Section \ref{nonparametric_PELT} we introduce our new efficient nonparametric search approach, nonparametric PELT (NP-PELT) and show how we can substantially improve the computational cost of this method.  In Section 4 we demonstrate the performance of our method on simulated data sets comparing our method with NMCD.  Finally in Section 5 we include some simulations which analyse the performance of NMCD for different scenarios and then we show how a nonparametric cost function can be beneficial in situations where we do not know the underlying distribution of the data.  In order to demonstrate our method we use heart rate data recorded whilst an individual is running.  

\section{Nonparametric Changepoint Detection}
\label{background}
\subsection{Model}
\label{model}
The model that we refer to throughout this paper is as follows.  Assume that we have  data, $x_1,...,x_n \in \mathbb{R}$ , 
that have been ordered based on some covariate information such as time or position along a chromosome.  For $v\geq u$ we denote  $x_{u:v} = \{x_{u},...,x_v\}$.   Throughout
we let $m$ be the number of changepoints, and the positions be $\tau_1,\ldots,\tau_m$. Furthermore we assume that $\tau_i$ is an integer and that 
$0 = \tau _0 < \tau_1 < \tau_2 < ... < \tau_m < \tau_{m+1} = n$.  Thus our $m$ changepoints split the data into $m+1$ segments, with the $i$th segment containing $x_{\tau_{i-1}+1:\tau_{i}}$.

As in \cite{Zou2014} we will let $F_i(t)$ be the (unknown) cumulative distribution function (CDF) for the $i$th segment, and $\hat{F}_i(t)$ the empirial CDF. In other words
\begin{align}
\hat{F}_i(t) =\frac{1}{\tau_i-\tau_{i-1}} \times \left(\sum_{j=\tau_{i-1}+1}^{\tau_i} \mathbf{1}\{x_j < t\} + 0.5 \times \mathbf{1}\{x_j = t\} \right).
\end{align} 
Finally we let $\hat{F}(t)$ be the empirical CDF for the full data set.

\subsection{Nonparametric maximum likelihood}
\label{nonparametric_max}

If we have $n$ data points that are independent and identically distributed with CDF $F(t)$, then, for a fixed value of $t$, the empirical CDF will satisfy 
$n\hat{F}(t)\sim \mbox{Binomial}(n,F(t))$. Hence the log-likelihood of $F(t)$ is given by: $n\{\hat{F}(t) \log ({F}(t)) + (1 - \hat{F}(t)) \log (1-{F}(t))\}.$ 
This log-likelihood is maximised by the value of the empirical CDF, $\hat{F}(t)$. We can thus use minus the maximum value of
this log-likelihood as a segment cost function. So for segment $i$ we have a cost that is $-\mathcal{L}_{np}(x_{\tau_{i-1}+1:\tau_i};t)$ where
\begin{eqnarray}
\mathcal{L}_{np}(x_{\tau_{i-1}+1:\tau_i};t) = (\tau_{i} - \tau_{i-1})\times 
[\hat{F}_i(t) \log \hat{F}_i(t) + (1-\hat{F}_i(t)) \log (1-\hat{F}_i(t))]. \label{eqn:obj2}
\end{eqnarray}
We can then define a cost of a segmentation as the sum of the segment costs. Thus to segment the data with $m$ changepoints we minimise $-\sum_{i=1}^{m+1}\mathcal{L}_{np}(x_{\tau_{i-1}+1:\tau_i};t)$.

\subsection{Nonparametric multiple changepoint detection}
\label{nmcd}

One problem with the segment cost as defined by (\ref{eqn:obj2}) is that it only uses information about the CDF evaluated at one value of $t$ and that the choice of $t$ can have detrimental effects on the resulting segmentations.  
To overcome this \cite{Zou2014} suggest defining a segment cost which integrates (\ref{eqn:obj2}) over different values of $t$. They suggest a cost function for
a segment with data $x_{u:v}$ that is
\begin{equation} \label{eq:intcost}
 \int_{-\infty}^{\infty} -\mathcal{L}_{np}(x_{u:v};t) dw(t),
\end{equation}
with a weight, $dw(t) = \{{F}(t)(1-{F}(t))\}^{-1} d{F}(t)$, that depends on the CDF of the full data.  This weight is chosen to produce a powerful goodness of fit test \citep{Zhang2002}.  As this is unknown they approximate it by the empirical CDF of the full data, and then further approximate the integral by a
sum over the data points. This gives the following objective function
\begin{align}
 Q_{\text{NMCD}}(\tau_{1:m};x_{1:n})= -n\sum_{i=1}^{m+1} \sum_{t = 1}^{n}(\tau_{i} - \tau_{i-1})
\times \frac{\hat{F}_i(t) \log \hat{F}_i(t) + (1-\hat{F}_i(t))\log(1-\hat{F}_i(t))}{(t - 0.5)(n-t+0.5)}.\label{eqn:obj}
\end{align}
For a fixed $m$ this objective function is minimised to find the optimal segmentation of the data.

In practice a suitable choice of $m$ is unknown, and \cite{Zou2014} suggest estimating $m$
using the Schwarz' Information criterion \citep{Schwarz1978}. That is, they minimise
\begin{align}
\text{SIC} = \min_{m;\tau_1,...,\tau_m} \left\{Q_{\text{NMCD}}(\tau_{1:m};x_{1:n}) + m \xi_n\right\}, \label{eqn:penalty}
\end{align}
where $\xi_n$ is a sequence going to infinity. 

\subsection{NMCD Algorithm}

To maximise the objective function (\ref{eqn:obj}), \cite{Zou2014} use the dynamic programming algorithm Segment Neighbourhood Search \citep{IVANE.AUGERandCHARLESE.LAWRENCE1989}.  
This algorithm calculates the optimal segmentations, given a cost function, for each value of $m=1,\ldots,M$, where $M$ is a specified maximum number of changepoints to search for. 
If all the segment costs have been pre-computed then Segment Neighbourhood search has a computational cost of $\mathcal{O}(Mn^2)$.  However for NMCD the segment cost involves calculating 
\[
\sum_{t = 1}^{n}\frac{\hat{F}_i(t) \log \hat{F}_i(t) + (1-\hat{F}_i(t))\log(1-\hat{F}_i(t))}{(t - 0.5)(n-t+0.5)},
\]
and thus calculating the cost for a single segment is $O(n)$. Hence the cost of precomputing all segment costs is $O(n^3)$, and the resulting algorithm has a cost that is $O(Mn^2+n^3)$.
 
To reduce the computational burden when we have long data series, \cite{Zou2014} propose a screening step. They consider overlapping windows of length $2N_I$ for some $N_I \in \mathbb{R}$. For each
window they calculate the Cram\'{e}r-von Mises (CvM) statistic for a changepoint at the centre of the window. They then compare these CvM statistics, each corresponding to a different 
changepoint location, and remove a location as a candidate changepoint if its CvM statistic is smaller than any of the CvM statistics for locations within $N_I$ of it.
The number of remaining candidate changepoint positions is normally much smaller than $n$ 
and thus the computational complexity can be substantially  reduced.   
The choice of $N_I$ is obviously important, with larger values leading to the removal of more putative changepoint locations, but at the risk or removing true changepoint locations. In particular,
the rationale for the method is based on $N_I$ being smaller than any segment that you wish to detect.
As a default, \cite{Zou2014} 
recommend choosing $N_I = \lceil (\log n)^{3/2}/2 \rceil$ where $\lceil x \rceil$ denotes the smallest integer which is larger than $x$. 

\section{NP-PELT}
\label{nonparametric_PELT}

Here we develop a new, computationally efficient, way to segment data using a cost function based on (\ref{eq:intcost}). This involves firstly an alternative numerical approximation to the integral (\ref{eq:intcost}), which
is more efficient to calculate. In addition we use a more efficient dynamic programming algorithm, PELT \citep{Killick2012}, to then minimise the cost function.

\subsection{Improved Segment Cost}\label{section:improved}
To reduce the cost of calculating the segment cost, we approximate the integral by a sum with $K<<n$ terms. The integral in (\ref{eq:intcost}) involves a weight, 
and we first make a change of variables to remove this weight. 
\begin{lemma} \label{lem:1}
Let $c=-\log(2n-1)$. For $x \in [-1,1]$ define $p(x)=(1+\exp\{cx\})^{-1}$. Then
\begin{align}
\int_{\frac{1}{2n}}^{\frac{2n-1}{2n}} \mathcal{L}_{np}(x_{u:v};t) \{F(t)(1-F(t))\}^{-1} dF(t) 
=-c \int_{-1}^1 \mathcal{L}_{np}(x_{u:v};F^{-1}(p(x))) dx. \label{eq:lem}
\end{align} 
\label{theorem:NPPELT}
\end{lemma}
\begin{proof}
This follows from making the change of variable $F(t)=p(x)$.
 \end{proof}
Using Lemma \ref{theorem:NPPELT}, we suggest the following approximation, based on an approximation of (\ref{eq:lem}) using $K$ evenly spaced $x$-values. 
Fix $K$, and define $\gamma$ such that $K=c/\gamma$, with $c$ defined as in Lemma \ref{lem:1}. Let $t_1,\ldots,t_K$ be such that
$t_k$ is the $(1+(2n-1)\exp\{\gamma(2k-1)\})^{-1}$ empirical quantile of the data, then we approximate (\ref{eq:intcost}) by
\begin{equation} \label{eq:sumapprox}
\mathcal{C}_K(x_{u:v})= \frac{-2c}{K} \sum_{k=1}^K  \mathcal{L}_{np}(x_{u:v};t_k).
\end{equation}

The idea is that (\ref{eq:intcost}) gives higher weight to values of $t$ in the tail of the distribution of the data. Our approximation achieves this through a sum where each term
has equal weight, but where the $t_k$ values we choose are prefentially chosen from the tail of the distribution. 

The cost now for calculating the segment costs is $\mathcal{O}(K)$. If we choose $K=\lceil c/\gamma \rceil$, where $c$ is defined in Lemma \ref{lem:1}, for some fixed $\gamma$ then this will be $O(\log n)$.
We investigate the choice of $K$ empirically in Section \ref{Results}.

\subsection{Use of PELT}

We now turn to consider how the PELT approach of \cite{Killick2012} can be incorporated within this framework.  The PELT dynamic programming algorithm is able to solve minimisation problems of the form
\[
 Q_{\text{PELT}}(x_{1:n}) = \min_{m,\tau_{1:m}} \left \{\sum_{i=1}^{m+1}[ \mathcal{C}_K(x_{\tau_{i-1}+1:\tau_i}) + \xi_n] \right\}.
\]
It jointly minimises over both the number and position of the changepoints, but requires the prior choice of $\xi_n$, the penalty value for adding a changepoint. The PELT algorithm 
uses the fact that $Q_{\text{PELT}}(x_{1:n})$ is the solution of the recursion, for $v>1$
\begin{equation} \label{eq:rec}
 Q_{\text{PELT}}(x_{1:v})=\min_{u< v}\left( Q_{\text{PELT}}(x_{1:u})+ \mathcal{C}_K(x_{u+1:v}) + \xi_n\right).
\end{equation}
The interpretation of this is that the term in the brackets on the right-hand side of Equation \ref{eq:rec} is the cost for segmenting $x_{1:v}$ with the most recent changepoint at $u$. We then optimise over the location of this most recent changepoint.
Solving the resulting set of recursions leads to an $O(n^2)$ algorithm \citep{Jackson2005}, as (\ref{eq:rec}) needs to be solved for $v=2,\ldots,n$; and solving (\ref{eq:rec}) for a given value of $v$ involves
a minimisation over $v$ terms.

The idea of PELT is that we can substantially speed up solving (\ref{eq:rec}) for a given $v$ by reducing the set of values of $u$ we have to minimise over. This can be done through a simple rule that enables us to
detect time points $u$ which can never be the optimal location of the most recent changepoint at any subsequent time. For our application this comes from the following result
\begin{theorem}
 If at time $v$, we have $u<v$ such that
\begin{align} \label{eq:PELT}
Q_{\text{PELT}}(x_{1:u}) + \mathcal{C}_K(x_{u+1:v}) \geq Q_{\text{PELT}}(x_{1:v}),
\end{align}
then for any future time  $T > v$, $u$ can never be the time of the optimal last changepoint prior to $T$. 
\end{theorem}
\begin{proof}
This follows from Theorem 3.1 of Killick et al. (2012),  providing we can show that for any $u<v<T$
\begin{equation} \label{eqn:proof1}
 \mathcal{C}_K(x_{u+1:T}) \geq \mathcal{C}_K(x_{u+1:v})+\mathcal{C}_K(x_{v+1:T}).
\end{equation}

As $\mathcal{C}_K(\cdot)$ is a sum of $k$ terms, each of the form $-\mathcal{L}_{np}(\cdot;t_k)$ we need only show that for any $t$
\[
 \mathcal{L}_{np}(x_{u+1:T};t) \leq \mathcal{L}_{np}(x_{u+1:v};t)+\mathcal{L}_{np}(x_{v+1:T};t).
\]
Now if we introduce notation that $\hat{F}_{u,v}(t)$ is the empirical CDF for data $x_{u:v}$, we have 
\begin{align*}
\mathcal{L}_{np}(x_{u+1:T};t) &=(T-u)[\hat{F}_{u,T}(t)\log(\hat{F}_{u,T}(t))+(1-\hat{F}_{u,T}(t))\log(1-\hat{F}_{u,T}(t))] \\
 &= \{(v-u)[\hat{F}_{u,v}(t)\log(\hat{F}_{u,T}(t))+(1-\hat{F}_{u,v}(t))\log(1-\hat{F}_{u,T}(t)) ] \\
 &+ (T-v)[\hat{F}_{v,T}(t)\log(\hat{F}_{u,T}(t))+(1-\hat{F}_{v,T}(t))\log(1-\hat{F}_{u,T}(t))] \}\\
 &\leq \mathcal{L}_{np}(x_{u+1:v};t)+\mathcal{L}_{np}(x_{v+1:T};t),
\end{align*}
as required.
\end{proof}
Thus at each time-point we can check whether (\ref{eq:PELT}) holds, and if so prune time-point $u$. Under certain regularity conditions, \cite{Killick2012} show that
for models where the number of changepoints increases linearly with $n$, such substantial pruning occurs that the PELT algorithm will have an expected
computational cost that is $O(n)$. We call the resulting algorithm we obtain nonparametric PELT (NP-PELT).

\section{Results}
\label{Results}
\subsection{Performance of NMCD}\label{section:performance}

We firstly compare the NMCD algorithm with (NMCD+) and without screening (NMCD) using the {\texttt{nmcdr}} R package (\cite{nmcdr}), with the default choices for $N_I$ and $\xi_n$ detailed in Section \ref{nmcd}. 
We set up a similar simulation as in \cite{Zou2014}. That is, we simulate data from the following three models, where $J(x) = \{1+sgn(x)\}/2$.
\paragraph{Model 1:} $x_i = \sum_{j=1}^{M} h_j J(nt_i - \tau_j) + \sigma \xi_i$,  
where
\begin{align}
\{\tau_j/n\} &= \{0.1, 0.13, 0.15, 0.23, 0.25, 0.40, 0.44, 0.65, 0.76, 0.78, 0.81\}, \nonumber \\
\{h_j\} &= \{2.01, -2.51, 1.51, -2.01, 2.51, -2.11, 1.05, 2.16, -1.56, 2.56, -2.11\},\nonumber
\end{align}
and there are $n$ equally spaced $t_i$ in $[0,1]$. 
\paragraph{Model 2:} $x_i = \sum_{j=1}^{M} h_i J(nt_i - \tau_j) + \sigma \xi_i \prod_{j=1}^{\sum_{j=1}^{M}J(nt_i - \tau_j)} v_j$, 
where 
\begin{align}
\{\tau_j/n\} &= \{0.20, 0.40, 0.65, 0.85\}, \{h_j\} = \{3,0,-2,0\}, \; \text{and} \; \{v_j\} = \{1,5,1,0.25\}. \nonumber 
\end{align}
\paragraph{Model 3:} $x_i \sim F_j(x)$,
where $\tau_j/n = \{0.20, 0.50, 0.75\}$, $j = 1,2,3,4$,
and $F_1(x),...,F_4(x)$ corresponds to the standard normal, the standardized $\chi_{(3)}^2$ (with zero mean and unit variance), the standardized $\chi_{(1)}^2$ and the standard normal distribution respectively.  

The first model has $M=11$ changepoints, all of which are changes in location.  Model 2 has both changes in location and in scale and model 3 has changes in skewness and in kurtosis. 
For the first two models we also consider three distributions for the error, $\xi_i$: $N(0,1)$, Student's $t$ distribution with 3 degrees of freedom and the standardised chi-square distribution with one degree of 
freedom, $\chi_{(1)}^2$.  

To compare both the NMCD and NMCD+ we look at the proportion of true positive changepoints, $\xi({\mathbf{C}}||\hat{{\mathbf{C}}})$, and false positive changepoints, $\xi(\hat{\mathbf{C}}||{\mathbf{C}})$, i.e.,
\begin{align}
\xi({\mathbf{C}}||\hat{{\mathbf{C}}}) = \frac{\#\mathbf{C} \in \hat{\mathbf{C}}}{n_\mathbf{C}} \; \mbox{and} \; \xi(\hat{\mathbf{C}}||{\mathbf{C}}) = \frac{\#\hat{\mathbf{C}} \notin {\mathbf{C}}}{n_{\hat{\mathbf{C}}}},
\end{align}
where ${\hat{\mathbf{C}}}$ are the estimated changepoints; ${\mathbf{C}}$ are the true changepoints; $n_{\hat{\mathbf{C}}}$ is the length of the estimated changepoint set and $n_{{\mathbf{C}}}$ is the length of the true changepoint set. In addition we also compare the computational time taken to run both of these methods.  The results can be seen in Table \ref{table:NMCD}.

\begin{table*}[t]
\caption{Comparison of NCMD and NMCD+ methods.  Values in the table are mean (standard deviation in parentheses) for 100 replications.} \label{table:NMCD}
\vskip 0.15in
\begin{small}
\begin{sc}
\scalebox{0.7}{
\begin{tabular}{llllllllll}
\hline \\
&  \multicolumn{3}{c}{$\xi(\mathbf{C}||\hat{\mathbf{C}})$} &\multicolumn{3}{c}{$\xi(\hat{\mathbf{C}}||\mathbf{C})$} & \multicolumn{3}{c}{Time} \\
&&&&&&&&&\\
\hline \\
 (I)& NMCD+ & NMCD & NPPELT+ & NMCD+ & NMCD & NPPELT+ & NMCD+(s) & NMCD(min) & NPPPELT+(s)\\
 $ N(0,1)$ &0.91(0.08)&0.93(0.07)& 0.92(0.07) &0.09(0.08)  & 0.07(0.07) & 0.08(0.07) & 1.89(1.08) & 19.57(1.08) & 1.49(0.02)  \\
   $t_{(3)}$ & 0.76(0.13) & 0.81(0.12) & 0.79(0.12)& 0.24(0.13) & 0.23(0.12) & 0.21(0.12)  & 2.00(0.36)  & 20.91(0.36) & 1.62(0.07)  \\
   $\chi^2_{(3)}$ & 0.83(0.10) & 0.91(0.08) & 0.90(0.09) & 0.17(0.10) & 0.10(0.09) & 0.10(0.09)  & 2.43(0.26) & 19.4(0.26)& 1.57(0.02)\\
\hline \\
(II)\\$N(0,1)$ &0.39(0.17) &0.58(0.22) &0.57(0.22)& 0.63(0.17) &0.45(0.21) &0.43(0.22) &1.92(1.25)&19.38(1.24)&3.16(0.19) \\
   $t_{(3)}$ &0.33(0.16) &0.48(0.21) &0.50(0.21)
   & 0.69(0.16) &  0.57(0.20) &0.50(0.21) & 1.92(0.25) &19.77(0.25)& 3.92(0.43) \\
   $\chi^2_{(3)}$ &0.35(0.17) &0.49(0.21) &0.48(0.22)& 0.66(0.17) & 0.51(0.21) &0.52(0.22)  & 1.90(0.31) & 21.76(0.31)& 3.48(0.20) \\
\hline\\
(III)\\
  & 0.36(0.24) & 0.48(0.26)& 0.45(0.24) &0.64(0.24)  & 0.53(0.26) & 0.55(0.24)   & 0.06(0.28)  & 30.90(1.09) & 3.34(0.12)  \\
 \hline
\end{tabular}
}
\end{sc}
\end{small}
\vskip -0.1in
\end{table*}

It is clear from Table \ref{table:NMCD} that using the screening step (NMCD+) significantly improves the computational cost for this method.  
However using this screening step comes at a cost of not correctly detecting the true changepoints. It can be seen that in all cases NMCD+ detects fewer true positives and more false positives than NMCD.

\begin{figure}
\centering
    \begin{subfigure}[b]{0.3\textwidth}
	\includegraphics[width=\columnwidth, height = 5cm]{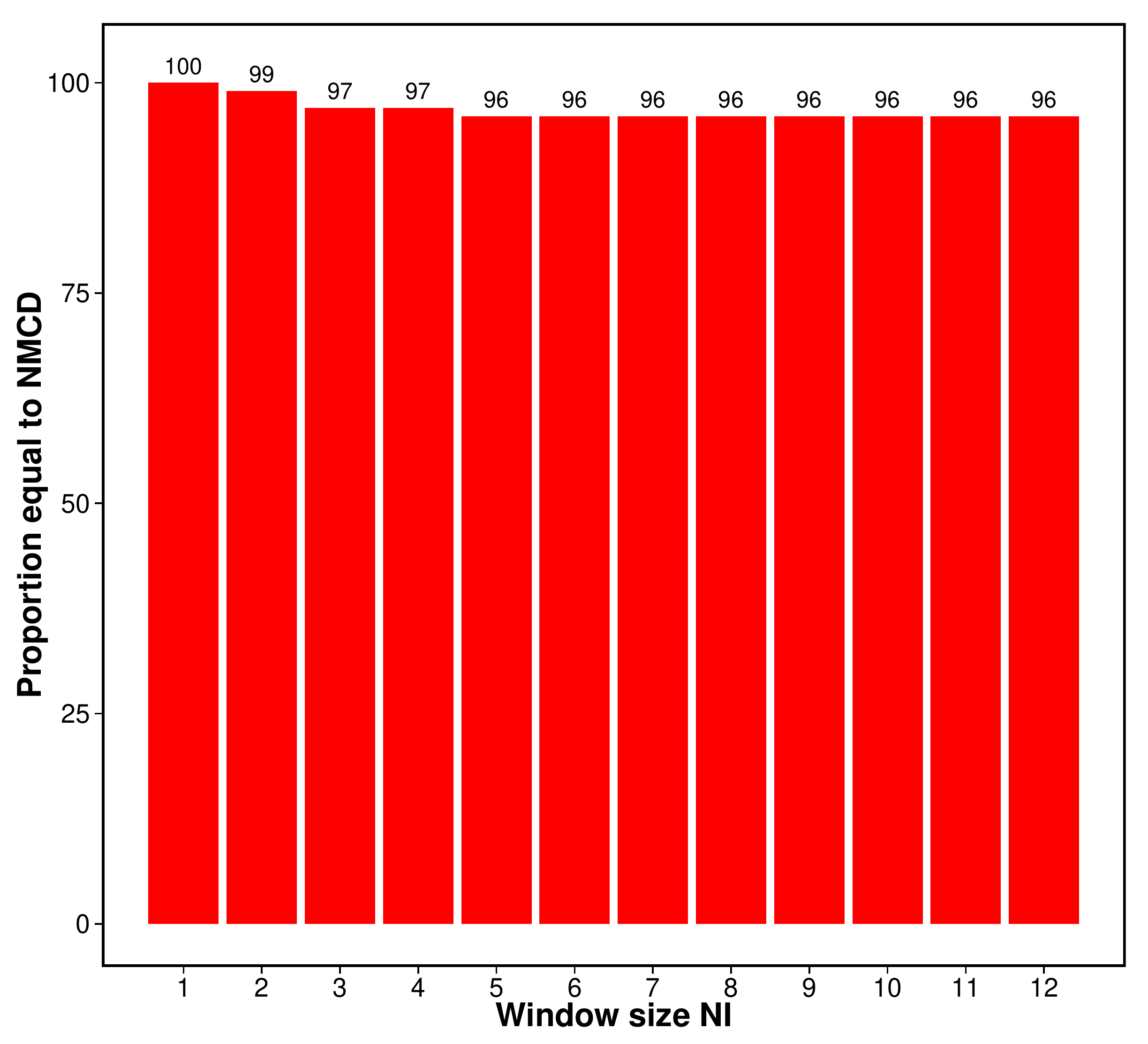}
	\caption{}
	\label{figure:barplot1_1}
\end{subfigure}
~
\begin{subfigure}[b]{0.3\textwidth}
\includegraphics[width=\columnwidth]{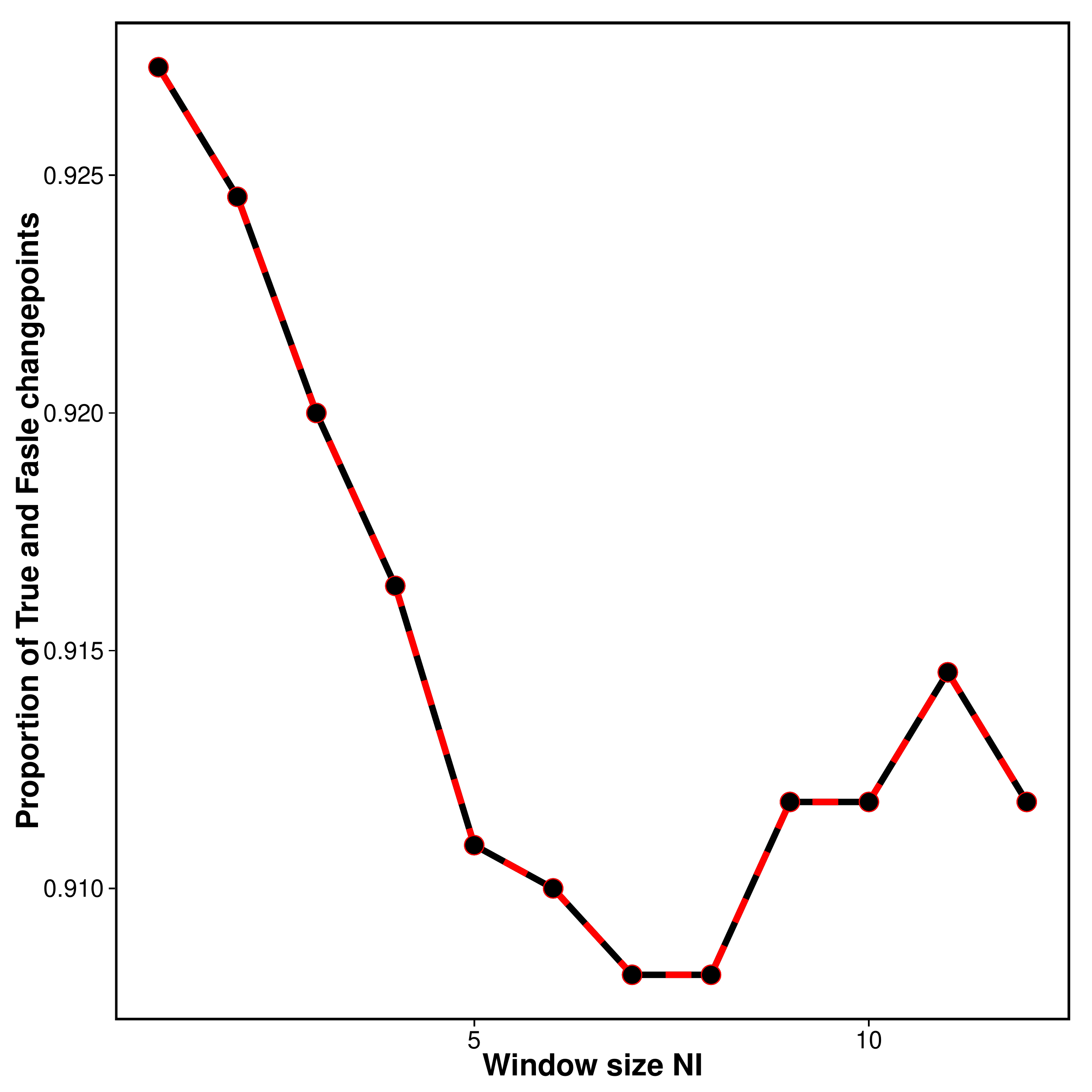}
\caption{}
\label{figure:True_false1_1}
\end{subfigure} 
~
\begin{subfigure}[b]{0.3\textwidth}
\includegraphics[width=\columnwidth]{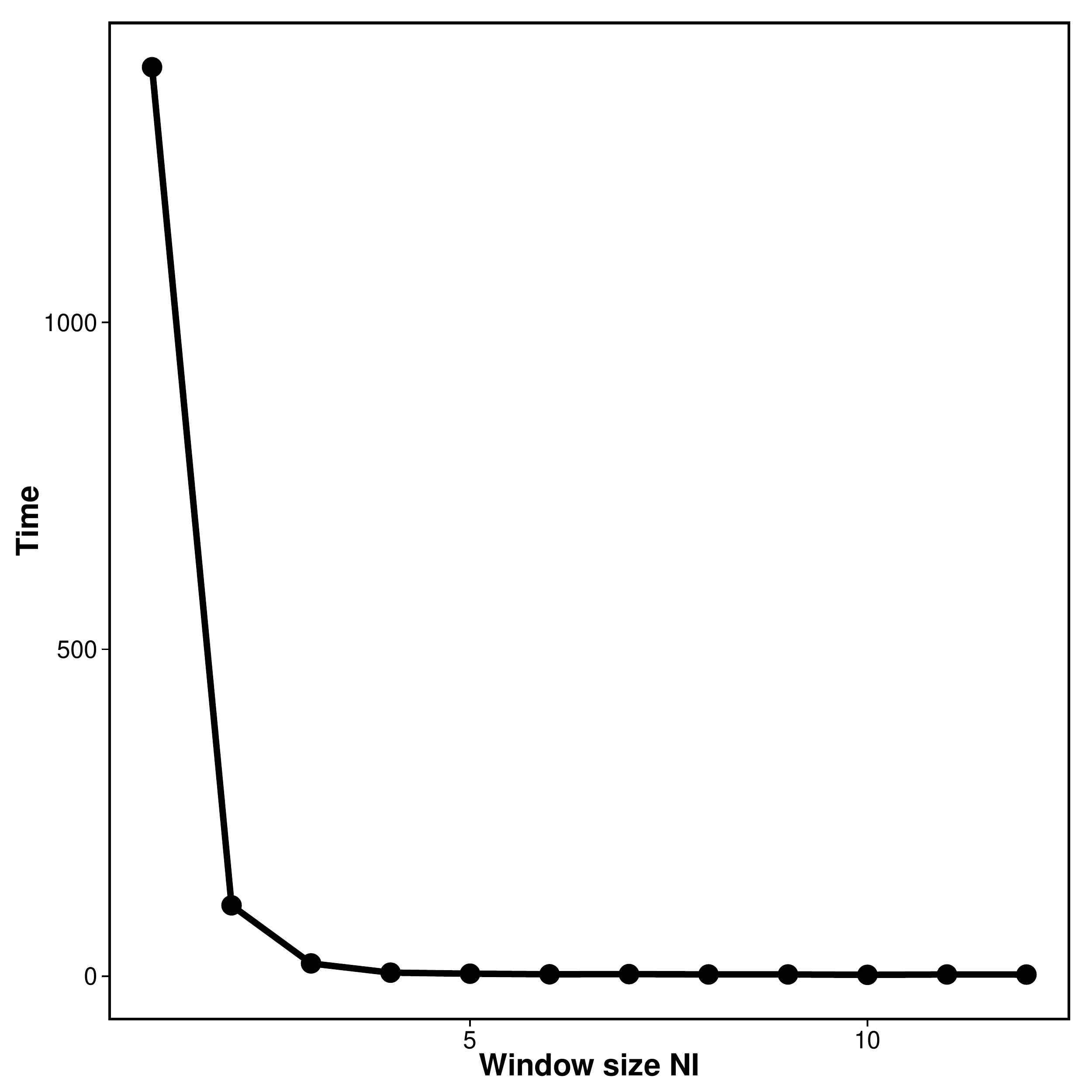}
\caption{}
\label{figure:Time1_1}
\end{subfigure} 
\caption{(a) The number of replications out of 100 in which using NMCD+ 	with varying $N_I$ results in the same results as NMCD without screening. (b) The proportion of true (black) and 1-False (red) changepoints detected with varying window size $N_I$. (c) The computational time (secs) for NMCD+ with increasing window size $N_I$.} 
\end{figure}

We now turn to consider the choice of the screening window $N_I$ further.  Using Model 1 with normal errors we can compare the results for different values of $N_I$. The default value for this data is $N_I=10$, but
we now repeat the analysis using $N_I \in \{1,\ldots,12\}$.  
Figure \ref{figure:barplot1_1} shows a bar plot of the number of times (in 100 simulations) that the window size resulted in the same changepoints as using NMCD without screening. 
Similarly Figure \ref{figure:True_false1_1} looks at the number of true and false positives found using the different window lengths in the screening step. Figure \ref{figure:Time1_1} shows the computational time taken for NMCD+ with varying window lengths $N_I$.  We found similar results for the other models.    

It is clear that whilst in the majority of the cases NMCD+ with the different $N_I$ gives the same result as NMCD but with an improved computational speed, there is a reasonable proportion of data sets where
NMCD+ does not find the optimal segmentation. As a result the segmentations it obtains are less accurate.

The NMCD method also requires us to choose a penalty value in order to pick the best segmentation. The default choice appears to work reasonably well,
but resulted in slight over-estimates of the number of changepoints for our three simulation scenarios.  These over-estimates suggest that the penalty value has been too small.

\subsection{Comparison to NMCD}\label{section:nmcd}
To compare NP-PELT to NMCD we initially used the same 3 models as above and again looked at the accuracy of the methods and the computational time.  
As before, to implement NMCD we used the \texttt{nmcdr} R package \cite{nmcdr} which has the bulk of the code in FORTRAN and we used R code to run NP-PELT. As R code is slower than FORTRAN, this means 
that the raw CPU timings we give will favour NMCD.

Firstly we investigate the speed up obtained by using the PELT pruning.  We implement PELT with the same segment costs as NMCD, and call this NP-PELT.  
We found that NP-PELT was at least 60 times faster than NMCD to run, but that NP-PELT is still an order of magnitude  slower than NMCD+. 

\subsection{Choice of $K$ in NP-PELT}
\begin{figure}[t]
\begin{subfigure}{0.45\textwidth}
\includegraphics[width=\columnwidth]{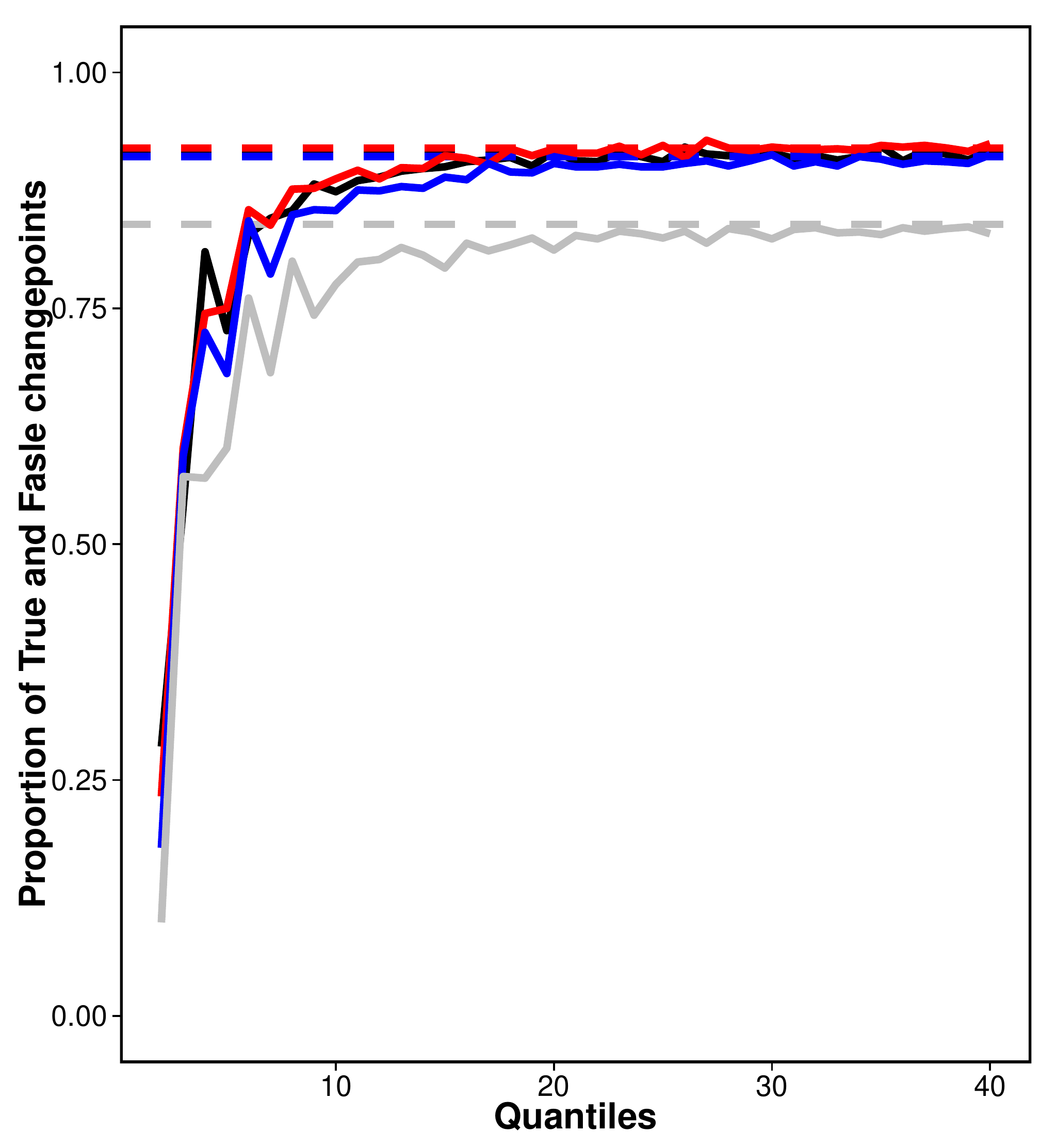}
\caption{}
\label{figure:K}
\end{subfigure} 
~
\begin{subfigure}{0.45\textwidth}
\includegraphics[width=\columnwidth, height = 8.2cm]{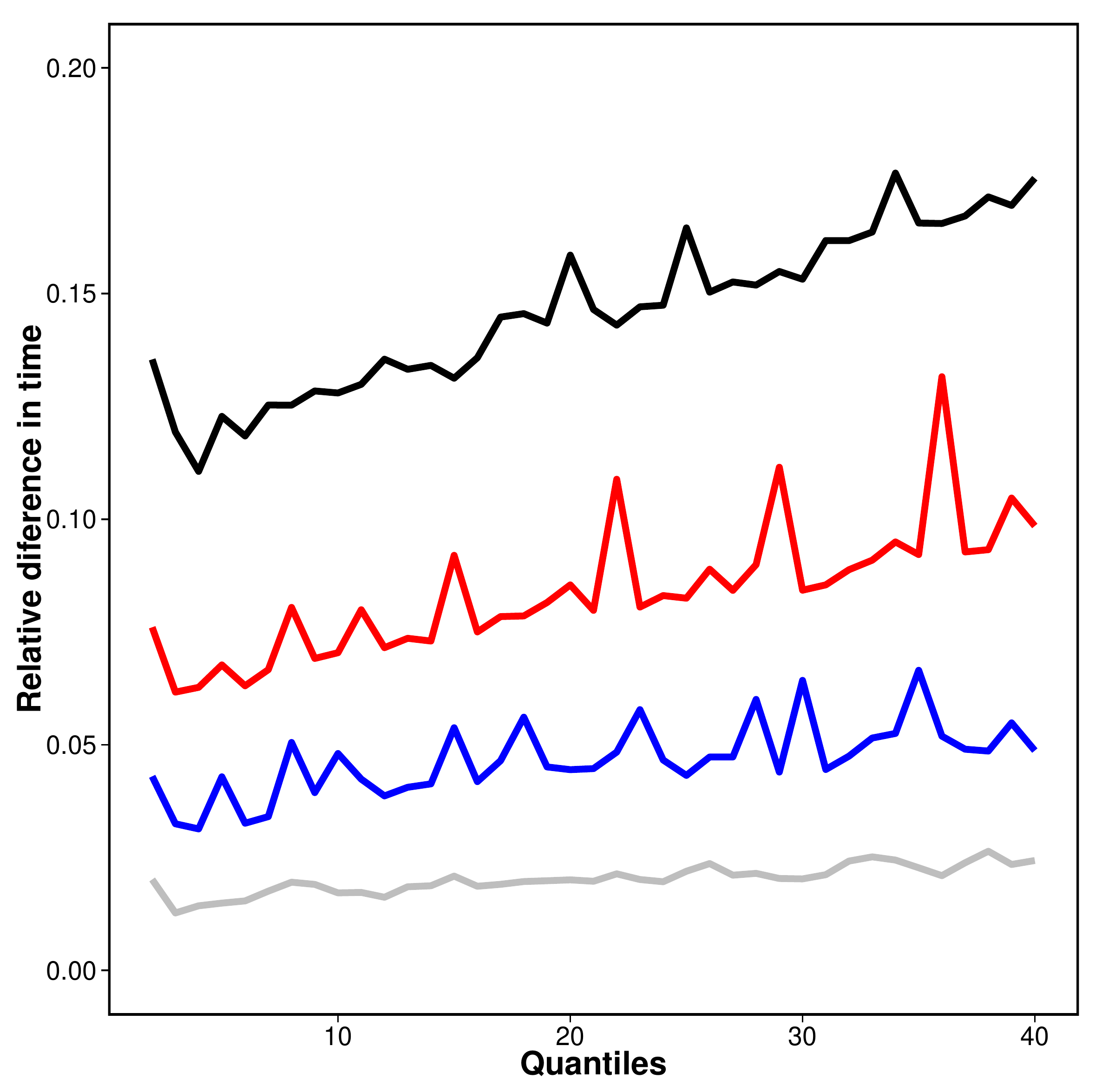}
\caption{}
\label{figure:time}
\end{subfigure}
\caption{(a) The proportion of true positive changepoints for a range of quantiles, $K$, in NP-PELT+ (solid) in comparison to NP-PELT (dashed).  Black: $n = 500$, red: $n = 1000$, blue: $n = 2000$ and grey: $n = 5000$.   (b) Relative speed of using NP-PELT+ compared to using NP-PELT. with varying number of quantiles, $K$.  Black: $n = 500$, red: $n = 1000$, blue: $n = 2000$ and grey: $n = 5000$.}   
\end{figure}

In order to use the improvement suggested in Section \ref{section:improved} for NP-PELT we first of all need to decide on an appropriate value for $K$. 
We use Model 1 again to assess the performance of NP-PELT using only $K$ quantiles of the data (NP-PELT+), for a range of values of $K$, in comparison to NP-PELT using the full data set.  
Here we only look at the model with normal errors and simulate data-series with lengths $n = (500,1000,2000,5000)$.  Further simulations using different error terms gave similar results. 
In order to assess performance we look at the proportion of true positives detected using both methods and also the computational cost.  Again we use 100 replications.  The results for the accuracy can be seen in Figure \ref{figure:K}. 

We can see from Figure \ref{figure:K} that as the number of quantiles increases the proportion of true change points detected using NP-PELT+ converges to the same result as NP-PELT.  
As the length of the data increases this convergence appears to happen more slowly, this can be seen from the grey lines in Figure \ref{figure:K}, which represent data of length 5000. 
We suggest using $K =\lceil 4\log(n) \rceil$ in order to conserve as much accuracy as possible. This choice corresponds to $K=25$, 28, 31 and 35 for $n = 500$, 1000, 2000 and 5000 respectively.

In addition to the accuracy we also look at the relative speed up of NP-PELT+ with various $K$ values in comparison to NP-PELT, i.e., 
\begin{align*}
\frac{\mbox{(speed of NP-PELT+)}}{\mbox{speed of NP-PELT}}.
\end{align*}
The results of this analysis can be seen in Figure \ref{figure:time}.  Clearly  as the number of quantiles increases the relative speed up decreases.  This is expected since the number of quantiles is converging to the whole data set which is used in NP-PELT.  We can also see that the relative speed up of NP-PELT+ increases with increasing data length.  

\subsection{Comparison of NMCD and NP-PELT+}

We next compare NP-PELT+ with $K = 4\log(n)$ to NMCD as above.   For this we perform an equivalent analysis to that of Section \ref{section:nmcd}.
The results for NP-PELT+ can be found in Table \ref{table:NMCD}.  In terms of accuracy we can see that NP-PELT+ is comparable to NMCD and is significantly faster to run.  
In comparison to NMCD+, in Models 2 and 3 NP-PELT+ is slightly slower.  This is to be expected however as the results of \cite{Killick2012} imply that
the cost of PELT will tend to be lower in situations with more changepoints.  In the first model there are 11 changepoints in data sets of length 1000,
however in Models 2 and 3 there are only 4 changepoints.  Despite this method being slower than NMCD+ it is more accurate.    

\section{Activity Tracking}\label{sec:AT} 

In this section we apply NP-PELT+ to try to detect changes in heart rate during a run.  Wearable activity trackers are becoming increasingly popular devices used to record step count, distances (based on the step count), sleep patterns and in some of the newer devices, such as the Fitbit change HR (Fitbit Inc., San Francisco, CA), heart rate.  The idea behind these devices is that the ability to monitor your activity should help you lead a fit and active lifestyle.  Changepoint detection can be used in daily activity tracking data to segment the day into periods of activity, rest and sleep.  

Similarly, many keen athletes, both professional and amateur, also use GPS sports watches which have the additional features of recording distance and speed which can be very beneficial in training, especially in sports such as running and cycling.  Heart rate monitoring during training can help make sure you are training hard enough without over training and burning out.  Heart rate is the number of heart beats per unit time, normally we express this as beats per minute (bpm).  

\subsection{Changepoints in heart rate data} 

In the changepoint and signal processing literature many authors have looked at heart rate monitoring in different scenarios (see for example \cite{Khalfa2012,Galway11,Billat20093798,Staudacher05}). 
\cite{Aubert03} give a detailed review of the influence of heart rate variability in athletes.  They highlight the difficulty of analysing heart rate measurements during exercise since no steady state is obtained due to the heart rate variability increasing according to the intensity of the exercise.  They note that one possible solution is to pre-process the data to remove the trend.

In this section we apply NP-PELT+ to see whether changes can be detected in the raw heart rate time series without having to initially pre-process the data.  We use a nonparametric approach since heart rate is a stochastic time dependent series and thus does not satisfy the conditions for an IID Normal model.  However we will compare the performance had we assumed that the data was Normal in Section \ref{sec:Normal}.   
The aim is to develop a method which can be used on data recorded from commercially available devices without the need to pre-process the data.      

\subsection{Range of Penalties} 

One disadvantage of NP-PELT+ over NMCD is that NP-PELT+ produces a single segmentation, which is optimal for the pre-chosen penalty value $\xi_n$. By comparison, NCMD finds a range of segmentations, one for each of 
$m=1,\ldots,M$ changepoints (though, in practice, the \texttt{nmcdr} package only outputs a single segmentation). Whilst there are default choices for $\xi_n$, these do not always work well, and there are advantages
to being able to compare segmentations with different number of changepoints.

\cite{Haynes14} propose a method, Changepoints over a Range Of PenaltieS (CROPS), which efficiently finds all the optimal segmentations for penalty values across a continuous range.  This involves an iterative
procedure which chooses values of $\xi_n$ to run NP-PELT on, based on the segmentations obtained from previous runs of NP-PELT for different penalty values. Assume we have a given range $[\xi_{\min},\xi_{\max}]$ for the penalty value,
and the optimal segmentations at $\xi_{\min}$ and $\xi_{\max}$ have $m_{\min}$ and $m_{\max}$ changepoints respectively. Then CROPS requires at most
$m_{\min}-m_{\max}+2$ runs of NP-PELT to be guaranteed to find all optimal segmentations for $\xi_n \in [\xi_{\min},\xi_{\max}]$. Furthermore, it is possible to recycle many of the calculations from
early runs of NP-PELT to speed up the later runs.

\subsubsection{Nonparametric Changepoint Detection}\label{sec:run}

\begin{figure}
\begin{center}
\includegraphics[width=\columnwidth]{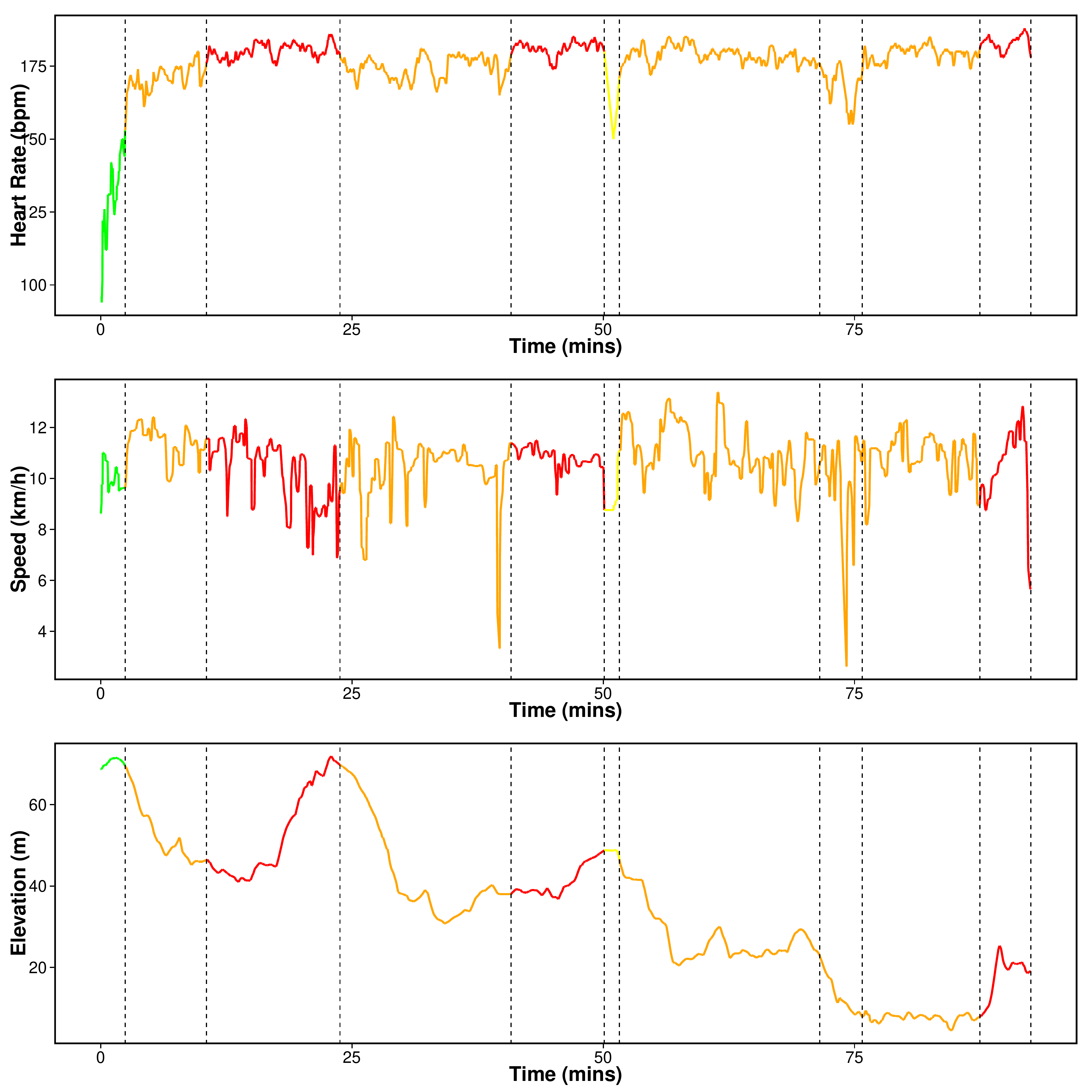}
\end{center}
\caption{Segmentations using NP-PELT+ with 10 changepoints.  We have colour coded the line based on the average heart-rate of each segment where red: peak, orange: anaerobic, yellow: aerobic and green: recovery.}
\label{figure:undulating}
\end{figure}

An example data set is given in Figure \ref{figure:undulating}, where we show heart-rate, speed and elevation recorded during a 10 mile run.  We will aim to segment this data using the heart-rate data only, but include the other two series in order that we may assess how well the segmentation of the heart-rate data relates to the obvious different phases of the run. In training many people use heart rate as an indicator of how hard they are working.  There are different heart rate zones that you can train in each of which enhances different aspects of your fitness \cite[]{BrainMac}. The training zones are defined in terms of percentages of a maximum heart-rate: peak (90-100$\%$), anaerobic (80-90$\%$), aerobic (70-80$\%$) and recovery ($<70\%$).  

\begin{figure}
\centering
    \begin{subfigure}[b]{0.3\textwidth}
    \caption{Elbow plot for NP-PELT+}
	\includegraphics[width=\columnwidth, height = 5cm]{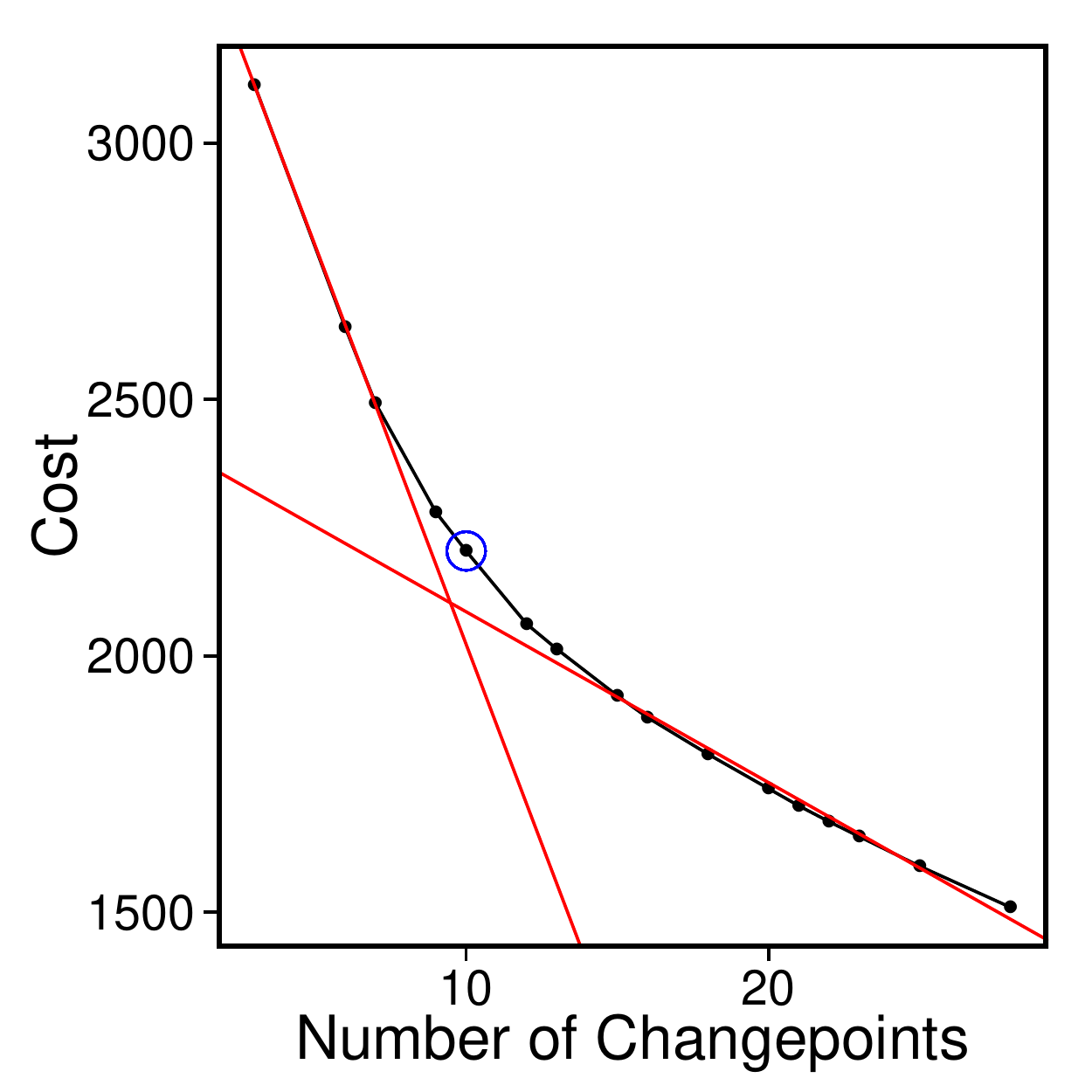}
	\label{figure:elbow1}
\end{subfigure}
~
\begin{subfigure}[b]{0.3\textwidth}
\caption{Elbow plot for a change in slope}
\includegraphics[width=\columnwidth]{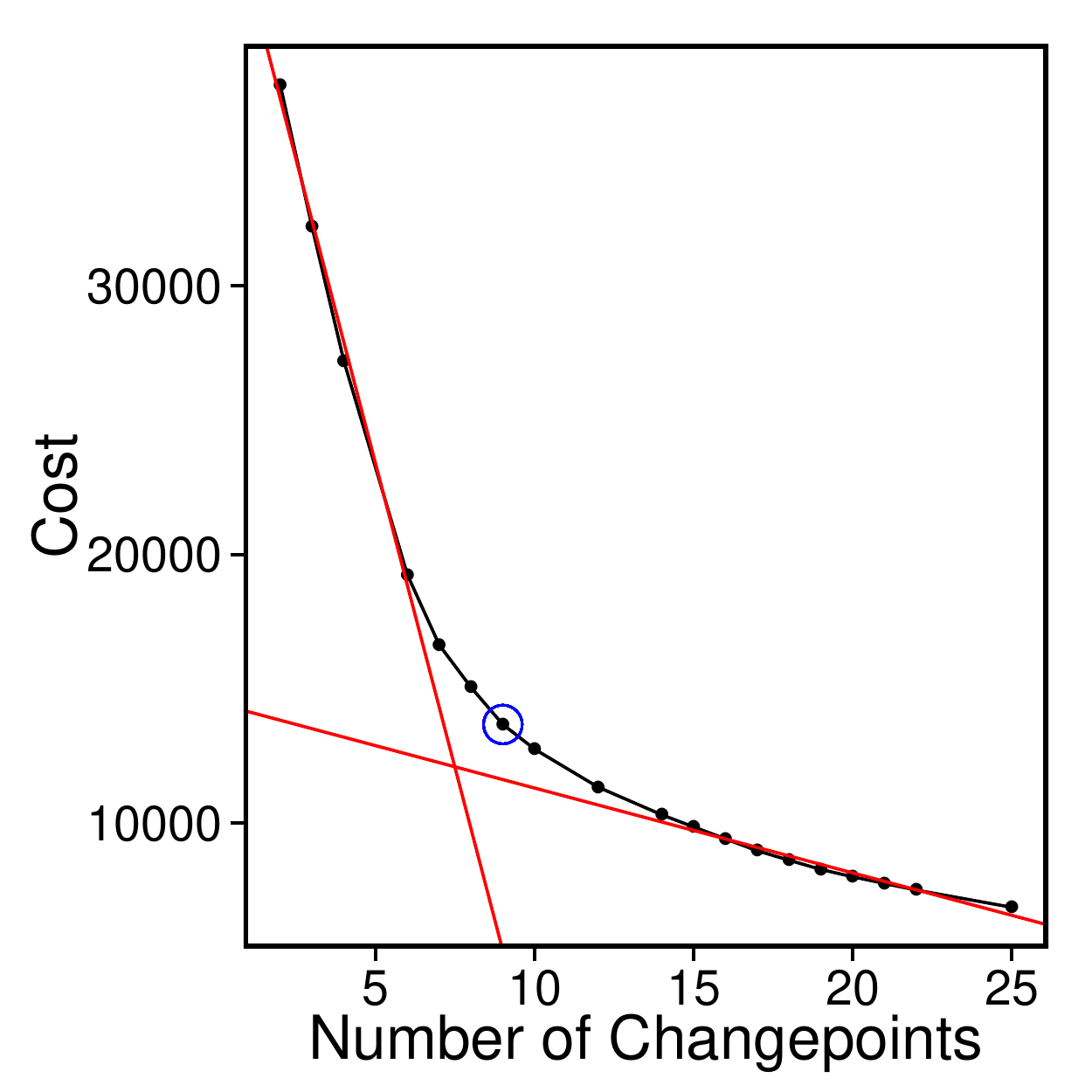}
\label{figure:elbow3}
\end{subfigure} 
\caption{The cost vs number of changepoints plotted for (a) NP-PELT+ and (b) Change in slope.  the red lines indicate the elbow and the blue circle highlights the point that we use as being the centre of the elbow. } \label{Fig:elbow}
\end{figure}

This example looks at detecting changes in heart rate over a long undulating run.  We use CROPS with NP-PELT+ with $\xi_{min} = 25$, $\xi_{max} = 200$ and $K = 4\log(n)$ (the results are similar for different $K$).  In order to choose the best segmentation we use the approach suggested by \cite{Lavielle2005}. This involves plotting the segmentation cost against the number of changepoints and then looking for an ``elbow'' in the plot.  The points on the ``elbow'' are then suggested to be the most feasible segmentations.  The intuition for this method is that as more true changepoints are detected the cost will decrease however as we detect more changepoints we are likely to be detecting false positives and as such the cost will not decrease as much.  The plot of the ``elbow'' for this example can be seen in Figure \ref{figure:elbow1}.  The elbow is not always obvious therefore the choice can be subjective however this gives us a method for roughly choosing the best segmentations which we can then explore further. We have highlighted the points on the ``elbow'' as the points which are between the two red lines.  

We decided from this plot that the segmentations with 9, 10, 12 and 13 changepoints are the best.  We illustrate the segmentation with 10 changepoints, the number of changepoints at the centre of the elbow in Figure \ref{figure:elbow1} indicated by the blue circle, in Figure \ref{figure:undulating}. The segments have been colour coded based on the average heart-rate in each segment.  That is red: peak, orange: anaerobic, yellow: aerobic and green: recovery.  Alternative segmentations from the number of changepoints on the elbow can be found in the supplementary material.  

We superimpose the changepoints detected in the heart rate onto the plots for speed and elevation to see if we can explain any of the changepoints. The first segment captures the ``warm-up'' where the heart-rate is on average in the recovery zone but is rising to the anaerobic zone. The heart-rate in the second segment is in the anaerobic zone but changes to the peak zone in segment three.  This change initially corresponds to an increase in speed and then it is because of the steep incline.  The third changepoint matches up to the top of the elevation which is the start of the fourth segment where the heart-rate drops into the anaerobic zone whilst running downhill.  The fifth segment is red which might be as a result of both the speed being slightly higher than the previous segment and consistent, and a slight incline in elevation. This is followed by a brief time in the aerobic zone which could be due to a drop in speed.  The heart-rate in the next three segments stays in the anaerobic zone.  The changepoints that split this section into three segments relate to the dip in speed around 75 minutes.  In the final segment the heart-rate is in the peak zone which corresponds to an increase in elevation and an increase in speed (a sprint finish).  

\subsection{Piece-wise linear model}\label{sec:Normal}
\begin{figure}
\begin{center}
\includegraphics[width=\columnwidth]{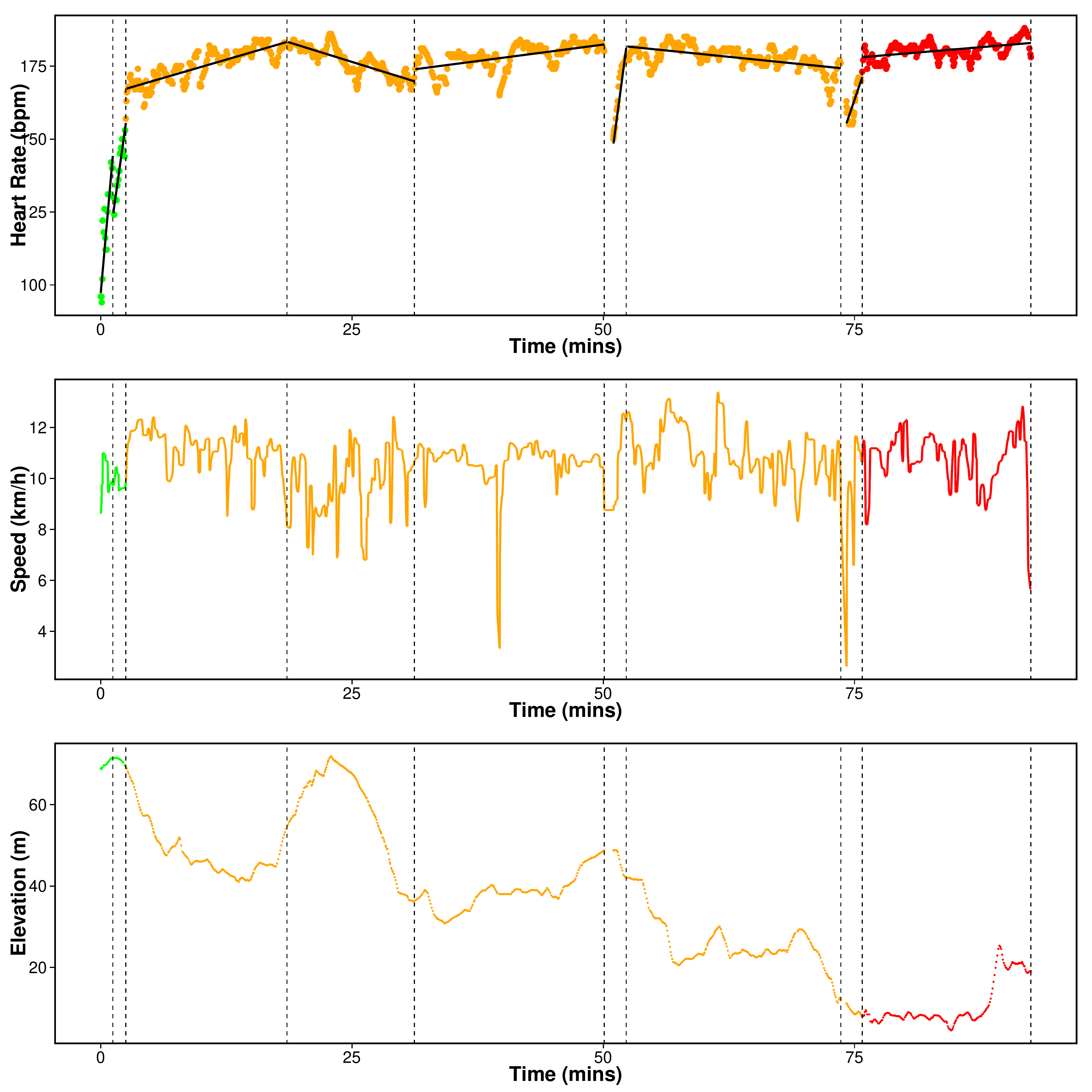}
\end{center}
\caption{Segmentations using change in slope with 9 changepoints.  We have colour coded the line based on the average heart-rate of each segment where red: peak, orange: anaerobic, yellow: aerobic and green: recovery. The solid black line in the top plot is the best fit for the mean within each segment.}
\label{figure:change_in_trend}
\end{figure}

For comparison we look at estimating the changepoints based on a penalised likelihood approach that assumes the data is normally distributed with a mean that is piecewise linear within each segment.
To find the best segmentation we use PELT with a segment cost proportional to minus the log-likelihood of our model:
\begin{align}
\mathcal{C}(y_{s:t})=min_{\theta_1,\theta_2} \sum_{u=s}^t (y_u-\theta_1-u\theta_2)^2,
\end{align}
where $\theta_1$ and $\theta_2$ and the estimates of the segment intercept and slope, respectively. 
We use CROPS to find the best segmentation under this criteria for a range of penalties.
The resulting elbow plot can be seen in Figure \ref{figure:elbow3}.  We can see that the number of changepoints for the feasible segmentations is similar to the number of changepoints for using NP-PELT+.  Figure \ref{figure:change_in_trend} shows the segmentation with 9 changepoints which we have deduced to being the number of changepoints in the centre of the elbow in Figure \ref{figure:elbow3}.  Alternative segmentations from the number of changepoints on the elbow can be found in the supplementary material.  \\

It is obvious from the first look at Figure \ref{figure:change_in_trend} that the change in slope method has not detected segments where the average heart-rate is different to the surrounding segments.  
The majority of the plot is coloured orange with only changes in the first and last segments.  The change in slope method splits the ``warm-up'' period into two segments whereas having this as 
one segment appears more appropriate.  Unlike NP-PELT+ the change in slope does not detect changes which correspond to the change in elevation and thus NP-PELT+ appears to split the heart-rate data into 
more appropriate segments which relate to different phases of the run.    

\section{Conclusion}
\label{Conclusion}

We have developed a new algorithm, NP-PELT, to detect changes in data series where we do not know the underlying distribution. 
This method is an adaption of the NMCD method proposed by \cite{Zou2014}. The main advantage of NP-PELT over NMCD is that it is orders of magnitude faster.  
We initially reduced the time to calculate the cost of a segment from $\mathcal{O}(n)$ to $\mathcal{O}(\log n)$ by simplifying the definition of the segment cost. 
\cite{Zou2014} use a screening step to improve the computational time and even though this is slightly faster than NP-PELT we show that it isn't as accurate.   
We have also shown that non-parametric changepoint detection, using NP-PELT, holds promise for segmenting data from activity trackers. We were able to segment heart-rate data recorded during a run into
meaningful segments that correspond to different phases of the run, and can be related to different regimes of heart-rate activity.

\textbf{Acknowledgements} 
Haynes gratefully acknowledges the support of the EPSRC funded EP/H023151/1 STOR-i centre for doctoral training and the Defence Science and Technology Laboratory.

\bibliographystyle{apalike}
\bibliography{example_paper}

\begin{thebibliography}{}

\bibitem[Aubert et~al., 2003]{Aubert03}
Aubert, A.~E., Seps, B., and Beckers, F. (2003).
\newblock Heart rate variability in athletes.
\newblock {\em Sports Medicine}, 33(12):889--919.

\bibitem[Aue and Horváth, 2013]{Aue2013}
Aue, A. and Horváth, L. (2013).
\newblock Structural breaks in time series.
\newblock {\em Journal of Time Series Analysis}, 34(1):1--16.

\bibitem[Auger and Lawrence, 1989]{IVANE.AUGERandCHARLESE.LAWRENCE1989}
Auger, I. and Lawrence, C. (1989).
\newblock {Algorithms for the Optimal Identification of Segment Neighborhoods}.
\newblock {\em Bulletin of Mathematical Biology}, 51(1):39--54.

\bibitem[Bai and Perron, 1998]{Bai1998}
Bai, J. and Perron, P. (1998).
\newblock Estimating and testing linear models with multiple structural
  changes.
\newblock {\em Econometrica}, 66:47--78.

\bibitem[Baron, 2000]{Baron2000}
Baron, M. (2000).
\newblock {Nonparametric adaptive change-point estimation and on-line
  detection}.
\newblock {\em Sequential Analysis}, 19:1--23.

\bibitem[Bhattacharyya and Johnson, 1968]{Bhattacharyya1968}
Bhattacharyya, G. and Johnson, R. (1968).
\newblock {Nonparametric tests for shift at an unknown time point}.
\newblock {\em The Annals of Mathematical Statistics}, 39(5):1731--1743.

\bibitem[Billat et~al., 2009]{Billat20093798}
Billat, V.~L., Mille-Hamard, L., Meyer, Y., and Wesfreid, E. (2009).
\newblock Detection of changes in the fractal scaling of heart rate and speed
  in a marathon race.
\newblock {\em Physica A: Statistical Mechanics and its Applications},
  388(18):3798 -- 3808.

\bibitem[BrainMacSportsCoach, 2015]{BrainMac}
BrainMacSportsCoach (2015).
\newblock Heart rate training zones.
\newblock \url{https://http://www.brianmac.co.uk/hrm1.htm}.

\bibitem[Carlstein, 1988]{Carlstein1988}
Carlstein, E. (1988).
\newblock {Nonparametric change-point estimation}.
\newblock {\em The Annals of Statistics}, 16(1):188--197.

\bibitem[Davis et~al., 2006]{Davis2006}
Davis, R.~A., Lee, T. C.~M., and Rodriguez-Yam, G.~A. (2006).
\newblock {Structural Break Estimation for Nonstationary Time Series Models}.
\newblock {\em Journal of the American Statistical Association},
  101(473):223--239.

\bibitem[Dette and Wied, 2015]{DetteWied2015}
Dette, H. and Wied, D. (2015).
\newblock Detecting relevant changes in time series models.
\newblock {\em Journal of the Royal Statistical Society: Series B (Statistical
  Methodology)}.

\bibitem[Dumbgen, 1991]{Dumbgen1991}
Dumbgen, L. (1991).
\newblock {The asymptotic behavior of some nonparametric change-point
  estimators}.
\newblock {\em The Annals of Statistics}, 19(3):1471--1495.

\bibitem[Frick et~al., 2014]{frick2014multiscale}
Frick, K., Munk, A., and Sieling, H. (2014).
\newblock Multiscale change point inference.
\newblock {\em Journal of the Royal Statistical Society: Series B (Statistical
  Methodology)}, 76(3):495--580.

\bibitem[Fryzlewicz, 2014]{WildBS}
Fryzlewicz, P. (2014).
\newblock Wild binary segmenation for multiple change-point detection.
\newblock {\em Ann. Statist.}, 42:2243--2281.

\bibitem[Galway et~al., 2011]{Galway11}
Galway, L., Zhang, S., Nugent, C., McClean, S., Finlay, D., and Scotney, B.
  (2011).
\newblock Utilizing wearable sensors to investigate the impact of everyday
  activities on heart rate.
\newblock In Abdulrazak, B., Giroux, S., Bouchard, B., Pigot, H., and Mokhtari,
  M., editors, {\em Toward Useful Services for Elderly and People with
  Disabilities}, volume 6719 of {\em Lecture Notes in Computer Science}, pages
  184--191. Springer Berlin Heidelberg.

\bibitem[Guarnaccia et~al., 2015]{Guarnaccia2015}
Guarnaccia, C., Quartieri, J., Tepedino, C., and Rodrigues, E.~R. (2015).
\newblock {An analysis of airport noise data using a non-homogeneous Poisson
  model with a change-point}.
\newblock {\em Applied Acoustics}, 91:33--39.

\bibitem[Haynes et~al., 2015]{Haynes14}
Haynes, K., Eckley, I.~A., and Fearnhead, P. (2015).
\newblock Computationally efficient changepoint detection for a range of
  penalties.
\newblock {\em Journal of Computational and Graphical Statistics \textit{(to
  appear)}}.

\bibitem[Jackson et~al., 2005]{Jackson2005}
Jackson, B., Scargle, J.~D., Barnes, D., Arabhi, S., Alt, A., Gioumousis, P.,
  Gwin, E., Sangtrakulcharoen, P., Tan, L., and Tsai, T.~T. (2005).
\newblock {An algorithm for optimal partitioning of data on an interval}.
\newblock {\em Signal Processing}, pages 1--4.

\bibitem[Jandhyala et~al., 2013]{Jandhyala2013}
Jandhyala, V., Fotopoulos, S., MacNeill, I., and Liu, P. (2013).
\newblock Inference for single and multiple change-points in time series.
\newblock {\em Journal of Time Series Analysis}, 34(4):423--446.

\bibitem[Khalfa et~al., 2012]{Khalfa2012}
Khalfa, N., Bertandm, P.~R., Boudet, G., Chamoux, A., and Billat, V. (2012).
\newblock Heart rate regulation processed through wavelet analysis and change
  detection: some case studies.
\newblock {\em Acta Biotheoretica}, 60:109 -- 29.

\bibitem[Killick et~al., 2012]{Killick2012}
Killick, R., Fearnhead, P., and Eckley, I.~A. (2012).
\newblock {Optimal detection of changepoints with a linear computational cost}.
\newblock {\em Journal of the American Statistical Association},
  107(500):1590--1598.

\bibitem[Lavielle, 2005]{Lavielle2005}
Lavielle, M. (2005).
\newblock {Using penalized contrasts for the change-point problem}.
\newblock {\em Signal Processing}, 85(8):1501--1510.

\bibitem[Lee, 1996]{Lee1996295}
Lee, C.-B. (1996).
\newblock Nonparametric multiple change-point estimators.
\newblock {\em Statistics and Probability Letters}, 27(4):295 -- 304.

\bibitem[Lu and Zhang, 2002]{Lu2002}
Lu, L. and Zhang, H.-j. (2002).
\newblock {Speaker Change Detection and Tracking in Real-Time News Broadcasting
  Analysis}.
\newblock {\em Proceedings of the tenth ACM international conference on
  multimedia}, pages 602--610.

\bibitem[Matteson and James, 2013]{Matteson2013}
Matteson, D.~S. and James, N.~A. (2013).
\newblock {A Nonparametric Approach for Multiple Change Point Analysis of
  Multivariate Data}.
\newblock 14853:1--29.

\bibitem[Nam et~al., 2014]{Nam2014}
Nam, C. F.~H., Aston, J. A.~D., Eckley, I.~A., and Killick, R. (2014).
\newblock {The Uncertainty of Storm Season Changes: Quantifying the Uncertainty
  of Autocovariance Changepoints}.
\newblock {\em Technometrics}, (January 2015):00--00.

\bibitem[Olshen et~al., 2004]{Olshen2004}
Olshen, A.~B., Venkatraman, E.~S., Lucito, R., and Wigler, M. (2004).
\newblock {Circular binary segmentation for the analysis of array-based DNA
  copy number data.}
\newblock {\em Biostatistics (Oxford, England)}, 5(4):557--72.

\bibitem[Page, 1954]{Page1954}
Page, E. (1954).
\newblock {Continuous inspection schemes}.
\newblock {\em Biometrika}, 41(1):100--115.

\bibitem[Pettitt, 1979]{Pettitt1979}
Pettitt, A. (1979).
\newblock {A non-parametric approach to the change-point problem}.
\newblock {\em Applied statistics}, 28(2):126--135.

\bibitem[Ross and Adams, 2012]{Ross2012}
Ross, G. and Adams, N.~M. (2012).
\newblock {Two Nonparametric Control Charts for Detecting Arbitrary
  Distribution Changes}.
\newblock {\em Journal of Quality Technology}, 44(2):102--116.

\bibitem[Schwarz, 1978]{Schwarz1978}
Schwarz, G. (1978).
\newblock {Estimating the Dimension of a Model}.
\newblock {\em The Annals of Statistics}, 6(2):461--464.

\bibitem[Scott and Knott, 1974]{Scott1974}
Scott, A. and Knott, M. (1974).
\newblock A cluster analysis method for grouping means in the analysis of
  variance.
\newblock {\em Biometrics}, 30:507--512.

\bibitem[Staudacher et~al., 2005]{Staudacher05}
Staudacher, M., Telser, S., Amann, A., Hinterhuber, H., and Ritsch-Marte, M.
  (2005).
\newblock A new method for change-point detection developed for on-line
  analysis of the heart beat variability during sleep.
\newblock {\em Physica A: Statistical Mechanics and its Applications},
  349(3):582--96.

\bibitem[Yao, 1988]{Yao1988}
Yao, Y.-C. (1988).
\newblock {Estimating the number of changepoints via Schwarz' Criterion}.
\newblock {\em Statistics \& Probability Letters}, 6(3):181--189.

\bibitem[Zhang, 2002]{Zhang2002}
Zhang, J. (2002).
\newblock {Powerful goodness-of-fit tests based on the likelihood ratio}.
\newblock {\em Journal of the Royal Statistical Society Series B},
  64(2):281--294.

\bibitem[Zhang and Siegmund, 2007]{Zhang2007b}
Zhang, N.~R. and Siegmund, D.~O. (2007).
\newblock {A modified Bayes information criterion with applications to the
  analysis of comparative genomic hybridization data.}
\newblock {\em Biometrics}, 63(1):22--32.

\bibitem[Zou et~al., 2014]{Zou2014}
Zou, C., Yin, G., Feng, L., and Wang, Z. (2014).
\newblock {Nonparametric maximum likelihood approach to multiple change-point
  problems}.
\newblock {\em The Annals of Statistics}, 42(3):970--1002.

\bibitem[Zou and Zhange, 2014]{nmcdr}
Zou, C. and Zhange, L. (2014).
\newblock {\em nmcdr: Non-parametric Multiple Change-Points Detection}.
\newblock R package version 0.3.0.

\end{thebibliography}
\end{document}


\begin{center}
{\large\bf SUPPLEMENTARY MATERIAL}
\end{center}
\begin{description}

\item[Further Results - NP-PELT+]
\end{description}
\begin{figure}[H]
\begin{center}
\includegraphics[width=\columnwidth]{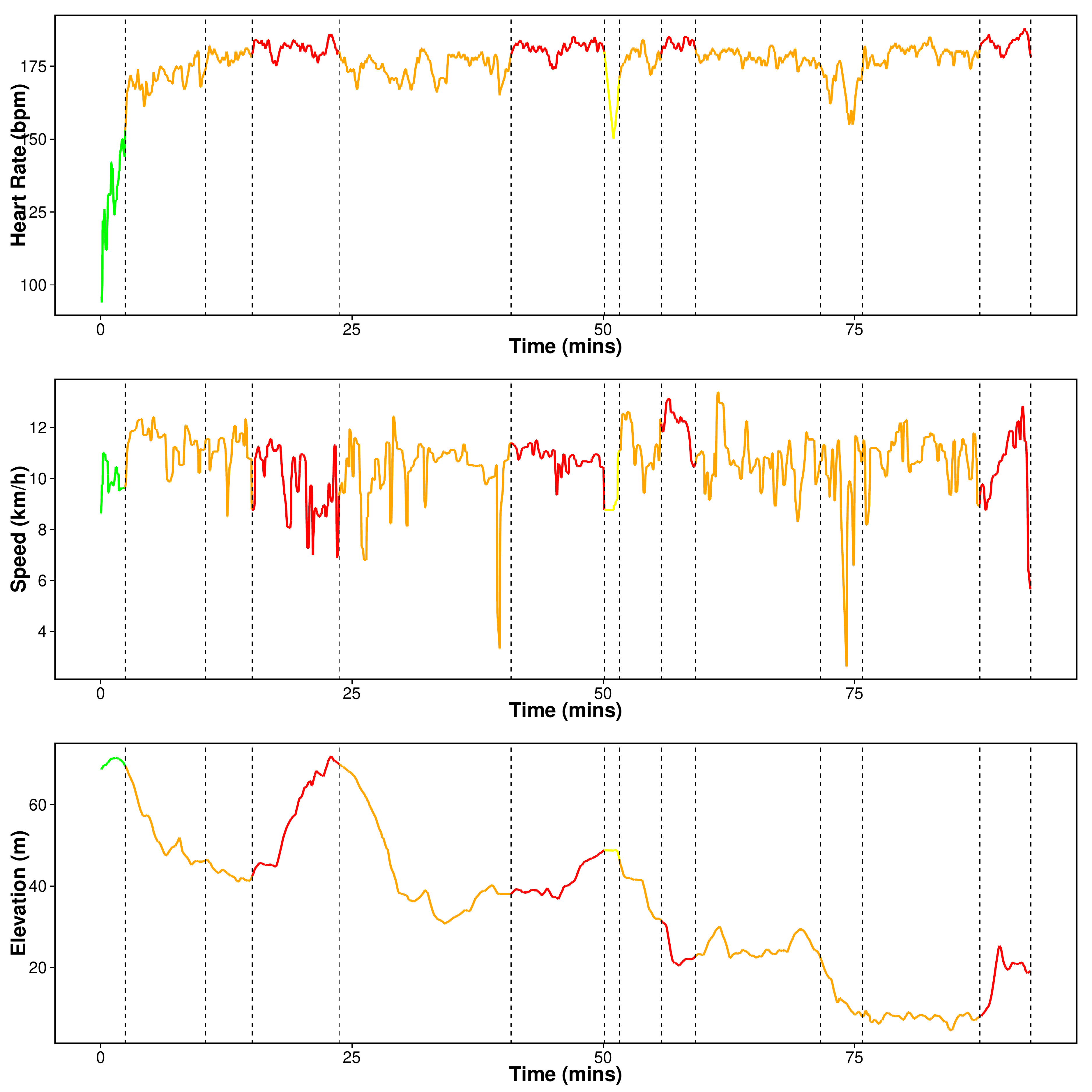}
\end{center}
\caption{Segmentations using NP-PELT+ with 13 changepoints.  We have colour coded the line based on the average heart-rate of each segment where red: peak, orange: anaerobic, yellow: aerobic and green: recovery.}
\label{figure:undulating1}
\end{figure}

\begin{figure}
\begin{center}
\includegraphics[width=\columnwidth]{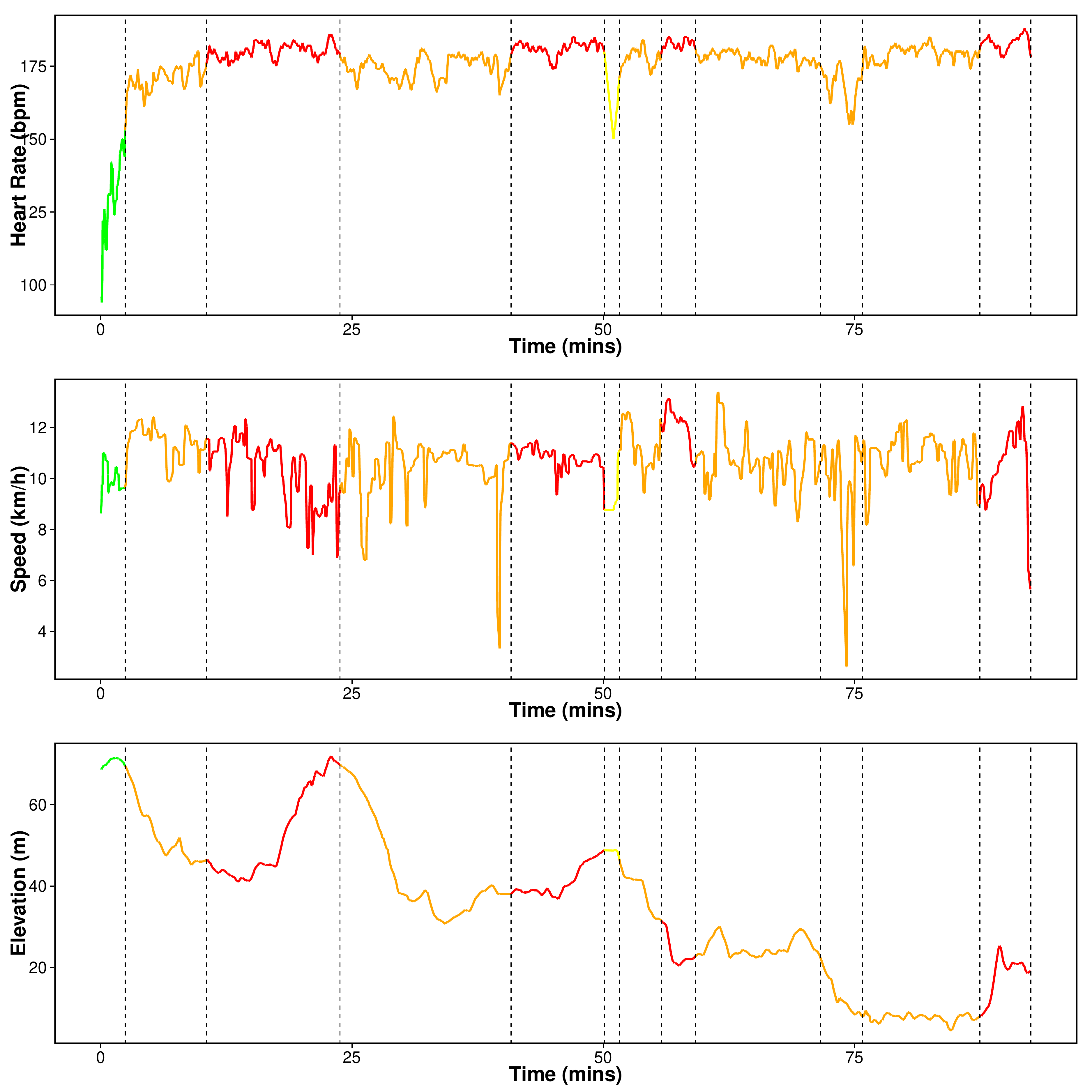}
\end{center}
\caption{Segmentations using NP-PELT+ with 12 changepoints.  We have colour coded the line based on the average heart-rate of each segment where red: peak, orange: anaerobic, yellow: aerobic and green: recovery.}
\label{figure:undulating2}
\end{figure}

\begin{figure}
\begin{center}
\includegraphics[width=\columnwidth]{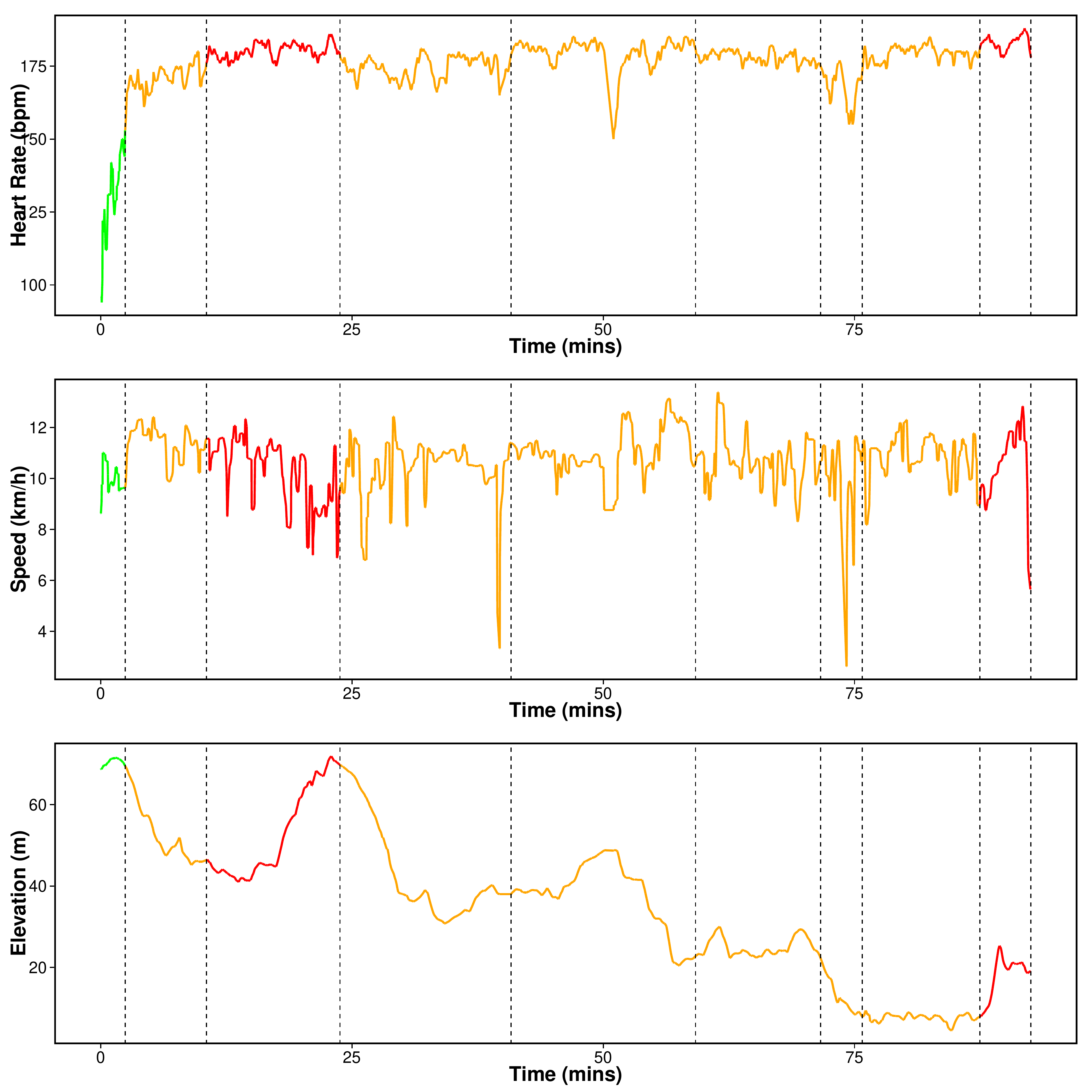}
\end{center}
\caption{Segmentations using NP-PELT+ with 9 changepoints.  We have colour coded the line based on the average heart-rate of each segment where red: peak, orange: anaerobic, yellow: aerobic and green: recovery.}
\label{figure:undulating3}
\end{figure}
 \clearpage
\begin{description}
\item[Further Results - Piece-wise linear]
\end{description}

\begin{figure}[H]
\begin{center}
\includegraphics[width=\columnwidth]{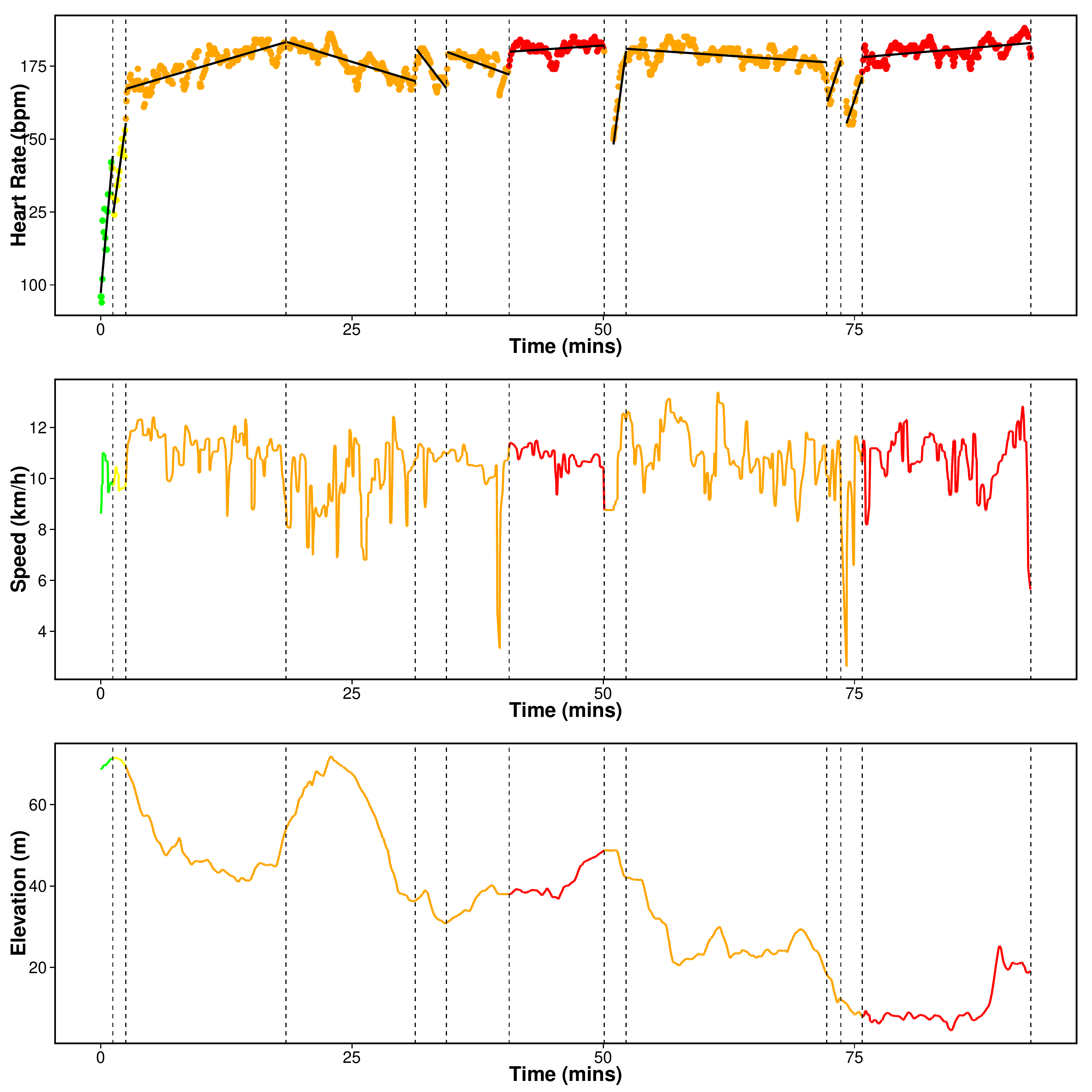}
\end{center}
\caption{Segmentations using change in slope with 12 changepoints.  We have colour coded the line based on the average heart-rate of each segment where red: peak, orange: anaerobic, yellow: aerobic and green: recovery. The solid black line in the top plot is the best fit for the mean within each segment.}
\label{figure:undulating4}
\end{figure}

\begin{figure}
\begin{center}
\includegraphics[width=\columnwidth]{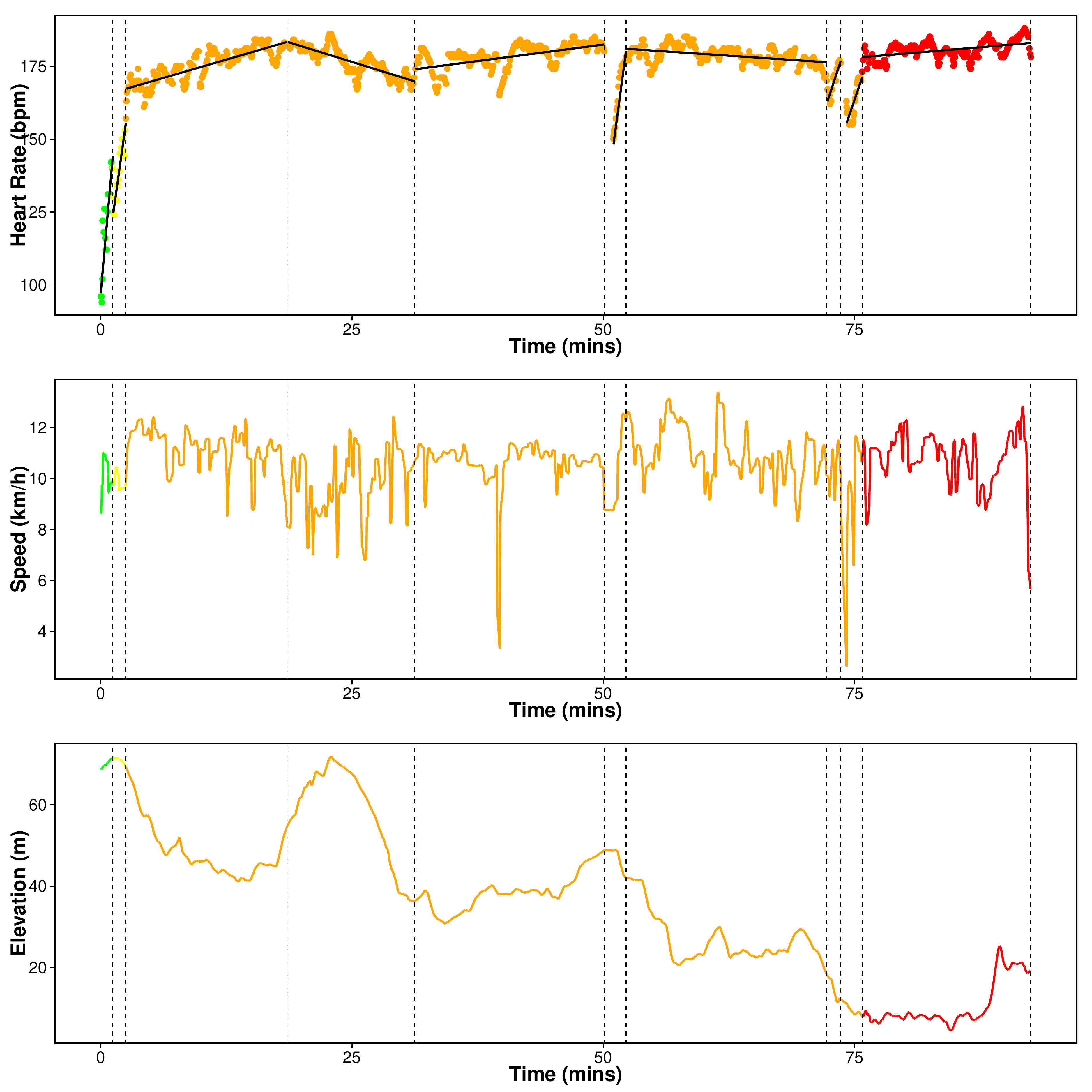}
\end{center}
\caption{Segmentations using change in slope with 10 changepoints.  We have colour coded the line based on the average heart-rate of each segment where red: peak, orange: anaerobic, yellow: aerobic and green: recovery. The solid black line in the top plot is the best fit for the mean within each segment.}
\label{figure:undulating5}
\end{figure}

\begin{figure}
\begin{center}
\includegraphics[width=\columnwidth]{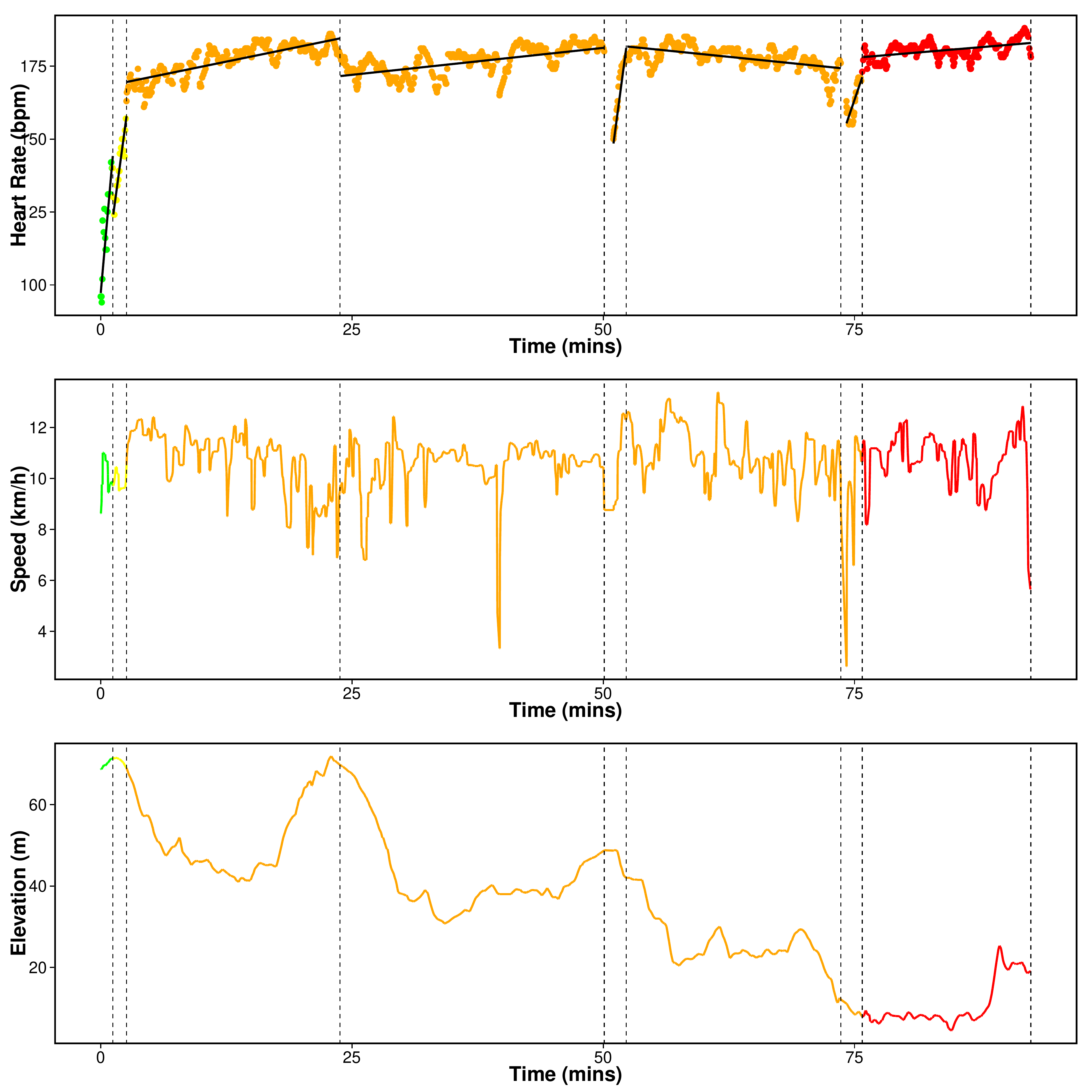}
\end{center}
\caption{Segmentations using change in slope with 8 changepoints.  We have colour coded the line based on the average heart-rate of each segment where red: peak, orange: anaerobic, yellow: aerobic and green: recovery. The solid black line in the top plot is the best fit for the mean within each segment.}
\label{figure:undulating6}
\end{figure}

\begin{figure}
\begin{center}
\includegraphics[width=\columnwidth]{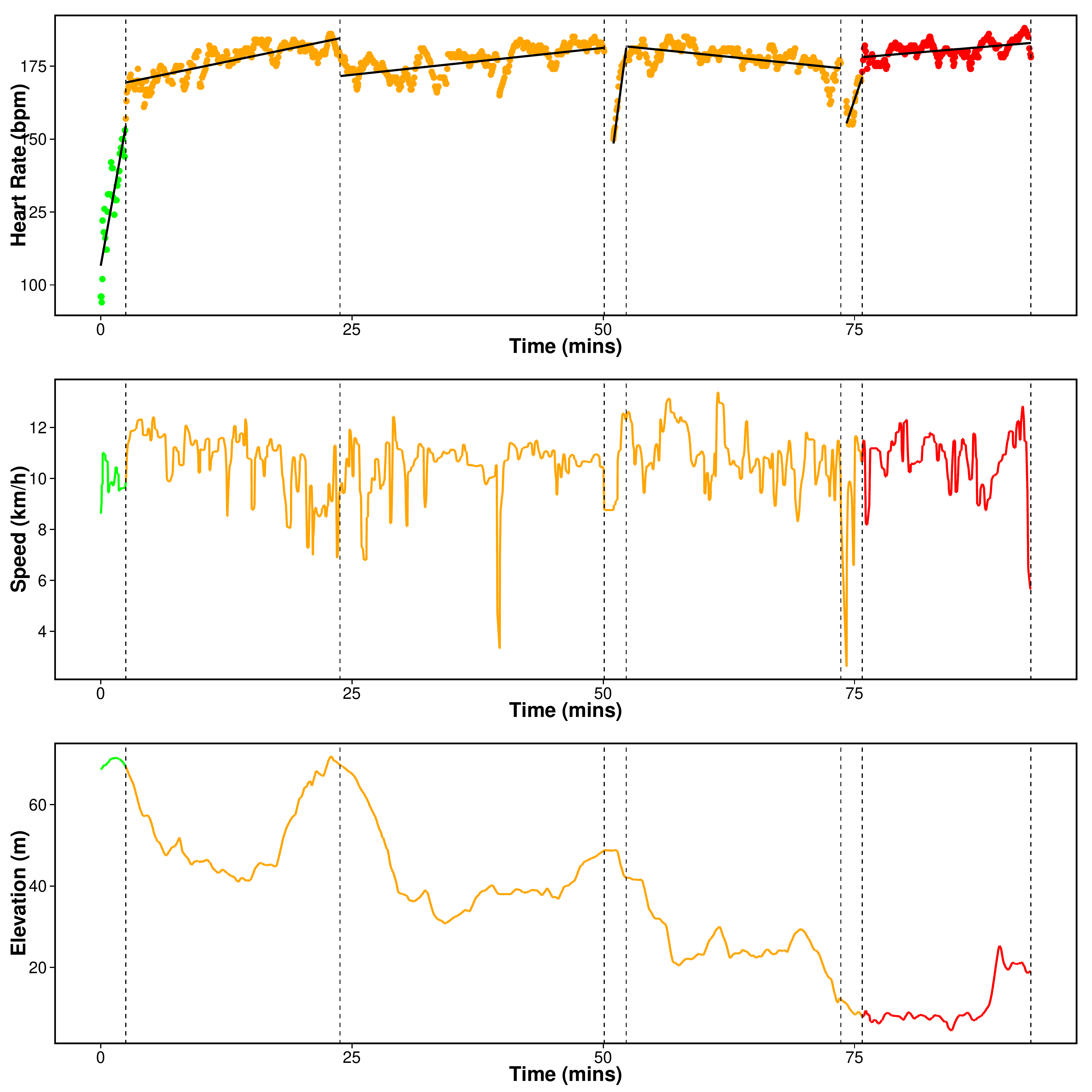}
\end{center}
\caption{Segmentations using change in slope with 7 changepoints.  We have colour coded the line based on the average heart-rate of each segment where red: peak, orange: anaerobic, yellow: aerobic and green: recovery. The solid black line in the top plot is the best fit for the mean within each segment.}
\label{figure:undulating7}
\end{figure}